\documentclass[fleqn,usenatbib]{mnras}

\usepackage{newtxtext,newtxmath}

\usepackage[T1]{fontenc}

\DeclareRobustCommand{\VAN}[3]{#2}
\let\VANthebibliography\thebibliography
\def\thebibliography{\DeclareRobustCommand{\VAN}[3]{##3}\VANthebibliography}

\usepackage{graphicx}	
\usepackage{amsmath}	
\usepackage{subfigure}

\newcommand{\pair}{\ensuremath{e^\pm}}

\newcommand{\msun}{\ensuremath{\,\rm{M}_\odot}}
\newcommand{\msunyr}{\ensuremath{\rm{M}_\odot\,\rm{yr}^{-1}}}
\newcommand{\mdot}{\ensuremath{\rm{\dot{M}}}}

\newcommand{\ergs}{\ensuremath{\rm{erg/s}}}

\newcommand{\mint}{\ensuremath{\rm{M_{init}}}}
\newcommand{\minit}{\ensuremath{\rm{M_{init}}}}

\newcommand{\mns}{\ensuremath{M_{\rm NS}}}

\newcommand{\yinit}{\ensuremath{\rm{Y_{init}}}}

\newcommand{\zinit}{\ensuremath{\rm{Z_{init}}}}

\newcommand{\logt}{\ensuremath{\rm{\log \left(T_{eff}/K\right)}}}
\newcommand{\logl}{\ensuremath{\rm{\log \left(L/L_{\odot}\right)}}}

\newcommand{\code}[1]{\texttt{#1}}

\newcommand{\MESA}{\code{MESA}}
\newcommand{\GYRE}{\code{GYRE}}

\newcommand{\kms}{{\mathrm{km\ s^{-1}}}}

\DeclareRobustCommand{\Eqref}[1]{Equation ~\ref{#1}}

\DeclareRobustCommand{\eqref}[1]{\Eqref{#1}}

\newcommand{\unitspace}{\ensuremath{\,}}
\newcommand{\usp}{\unitspace}

\newcommand{\unitstyle}[1]{\ensuremath{\mathrm{#1}}}
\newcommand{\power}[2]{\ensuremath{{#1}^{#2}}}

\newcommand{\centi}{\unitstyle{c}}
\newcommand{\kilo}{\unitstyle{k}}

\newcommand{\meter}{\unitstyle{m}}

\newcommand{\km}{\kilo\meter}   

\newcommand{\cm}{\centi\meter}
\newcommand{\gram}{\unitstyle{g}}

\newcommand{\grampercc}{\gram\usp\power{\cm}{-3}} 
\newcommand{\avgrhoc}{\ensuremath{\langle \rho_c \rangle}} 

\newcommand{\avgrhocgcm}{\ensuremath{\langle \rho_c / \grampercc \rangle}} 

\newcommand{\tzol}{Thorne–\.Zytkow objects}
\newcommand{\tzo}{T\.ZO}
\newcommand{\tzos}{T\.ZOs}
\newcommand{\tzosp}{T\.ZO's}

\newcommand{\medd}{\ensuremath{\dot{M}_{\rm Edd}}}
\newcommand{\ledd}{\ensuremath{L_{\rm Edd}}}
\newcommand{\emedd}{\ensuremath{\epsilon_{\dot{M}}}}
\newcommand{\eledd}{\ensuremath{\epsilon_{L}}}
\newcommand{\etanl}{\ensuremath{\eta_{\rm{\rm VL}}}}
\newcommand{\nisos}{399}
\newcommand{\lknee}{\ensuremath{L_{\rm{knee}}}}
\newcommand{\mlta}{\ensuremath{\rm{\alpha_{mlt}}}}

\newcommand{\nuclei}[2]{\ensuremath{\mathrm{^{#1}#2}}}

\newcommand{\helium}[1][4]{\nuclei{#1}{He}}
\newcommand{\lithium}[1][7]{\nuclei{#1}{Li}}
\newcommand{\beryllium}[1][9]{\nuclei{#1}{Be}}

\newcommand{\carbon}[1][12]{\nuclei{#1}{C}}
\newcommand{\nitrogen}[1][14]{\nuclei{#1}{N}}
\newcommand{\oxygen}[1][16]{\nuclei{#1}{O}}

\newcommand{\neon}[1][20]{\nuclei{#1}{Ne}}

\newcommand{\calcium}[1][40]{\nuclei{#1}{Ca}}
\newcommand{\scandium}[1][45]{\nuclei{#1}{Sc}}
\newcommand{\titanium}[1][48]{\nuclei{#1}{Ti}}

\newcommand{\iron}[1][56]{\nuclei{#1}{Fe}}

\newcommand{\nickel}[1][58]{\nuclei{#1}{Ni}}

\newcommand{\ruthenium}[1][102]{\nuclei{#1}{Ru}}

\defcitealias{cannon93}{C93}
\defcitealias{cannon92}{C92}

\title[\tzos]{Observational predictions for \tzol}

\author[R. Farmer et al]{
R. Farmer,$^{1}$\thanks{E-mail: rfarmer@mpa-garching.mpg.de}
M. Renzo,$^{2}$
Y. G\"{o}tberg,$^{3}\thanks{Hubble Fellow}$
E. Bellinger,$^{1,4}$
S. Justham,$^{1,5,6,7}$
and S.E. de Mink,$^{1,7}$
\\
$^{1}$Max-Planck-Institut für Astrophysik, Karl-Schwarzschild-Straße 1, 85741 Garching, Germany\\
$^{2}$Center for Computational Astrophysics, Flatiron Institute, New York, NY 10010, USA\\
$^{3}$The Observatories of the Carnegie Institution for Science, 813 Santa Barbara Street, Pasadena, CA 91101, USA\\
$^{4}$Stellar Astrophysics Centre, Aarhus University, 8000 Aarhus, Denmark\\
$^{5}$School of Astronomy \& Space Science, University of the Chinese Academy of Sciences, Beijing 100012, China\\
$^{6}$National Astronomical Observatories, Chinese Academy of Sciences, Beijing 100012, China\\
$^{7}$Anton Pannekoek Institute for Astronomy and GRAPPA, University of Amsterdam, NL-1090 GE Amsterdam, The Netherlands
}

\date{Accepted XXX. Received YYY; in original form ZZZ}

\pubyear{2023}

\begin{document}
\label{firstpage}
\pagerange{\pageref{firstpage}--\pageref{lastpage}}
\maketitle

\begin{abstract}

\tzol{} (\tzo) are potential end products of the merger of a neutron star with a non-degenerate star.
In this work, we have computed the first grid of evolutionary models of \tzos{} with the \MESA{} stellar evolution code.
With these models, we predict several observational properties of \tzos, including their surface temperatures and luminosities, pulsation periods, and nucleosynthetic products.
We expand the range of possible \tzo{} solutions to cover $3.45 \lesssim \logt \lesssim 3.65$ and $4.85 \lesssim \logl \lesssim 5.5$.
Due to the much higher densities our \tzos{} reach compared to previous models, if \tzos{} form we expect them to be stable over a larger mass range than previously predicted, without exhibiting a gap in their mass distribution.
Using the \GYRE{} stellar pulsation code we show that \tzos{} should have fundamental pulsation periods of 1000--2000 days, and period ratios of $\approx$0.2--0.3.
Models computed with a large \nisos{} isotope fully-coupled nuclear network show a nucleosynthetic signal that is different to previously predicted.
We propose a new nucleosynthetic signal to determine a star's status as a \tzo: the isotopologues $\titanium[44] \rm{O}_2$ and $\titanium[44] \rm{O}$, which will have a shift in their spectral features as compared to stable titanium-containing molecules.
We find that in the local Universe ($\sim$SMC metallicities and above) \tzos{} show little heavy metal enrichment, potentially explaining the difficulty in finding \tzos{} to-date.

\end{abstract}

\begin{keywords}
stars: evolution -- stars: abundances -- stars: interiors -- stars: variables
\end{keywords}



\section{Introduction}

\tzol{} ({\tzo}s) are the hypothetical unique product of the merger of a neutron star (NS) with a non-degenerate star leading to the formation of a single object \citep{thorne75,thorne77}. Depending on the mass of the combined star, it can be supported either via nuclear burning on/or near the surface of the NS and by accretion onto the NS \citep{eich89,biehle91}.

A number of potential \tzos{} candidates have been suggested: U Aqr \citep{vanture99}, HV 2112 \citep{levesque14}, HV 11417 \citep{beasor18}, and VX Sgr \citep{tabernero21}, though they are not without controversy \citep{coe98,vanture99,tout14,maccarone16,tabernero21}. Objects have been also been proposed that may form a \tzo{} (TIC 470710327 \citealt{eisner22}), or may be the remnants of a \tzo{} (1E161348-5055 \citealt{liu05}).
The issue with identifying a \tzo{} is distinguishing it from other similar stars, such as asymptotic giant branch (AGB) or super-AGB (SAGB) stars \citep{biehle91,biehle94,ogrady20,ogrady23}. \tzos{} are expected to appear as cool red supergiants (RSG) \citep{cannon92} (hereafter \citetalias{cannon92}), though many challenges remain in observing and modelling RSGs complicating their analysis \citep{levesque05,davies18}. Observations to date have relied on indirect measurements, such as their predicted unique nucleosynthetic signatures \citep{biehle94,paradijs95}. Future gravitational wave observations may provide an opportunity to detect either the formation of the \tzo{} \citep{nazin95,fraile23}, a rotating NS inside a \tzo{} \citep{demarchi21}, or a post-\tzo{} black hole (BH) \citep{cholis22}.

\tzos{} are expected to be fully convective down to a ``knee'' where it then transitions to a flat temperature profile until the material reaches the surface of the NS \citep{eich89}. At the knee, nuclear-processed material will be mixed outwards into the convective envelope, while fresh hydrogen is mixed into the region below the knee. Material may also be burnt at the base of the convective envelope. It is expected that the material is burnt via the interrupted rapid proton process (\texttt{irp}-process, \citealt{cannon93} hereafter \citetalias{cannon93}), where rapid proton captures \citep{wallace81,vanwromer94,fisker08} are interrupted when the material is mixed outwards by convection. This material then beta decays before being mixed back into the inner regions where additional proton captures can occur.

This \texttt{irp}-process can lead to the production of heavy elements such as Rubidium, Strontium, Yttrium, and Molybdenum \citepalias{cannon93}. It is also expected that the \tzos{} will be enriched in \calcium{} \citep{biehle91,biehle94}, which in normal stars is difficult to produce and mix to the surface \citep{tout14}. Lithium is also expected to be produced via \helium[3]($\alpha,\gamma$)\beryllium[7]($e^{-},\nu$)\lithium[7] \citep{cameron55}. In non-\tzo{} stars the \lithium[7] would be destroyed by high temperatures inside the star, but in a fully convective star the \beryllium[7] can be mixed outwards to cooler regions before it captures an electron \citep{Podsiadlowski95}. However, the nucleosynthetic signal is not unique, there are also \tzo{} imposters: SAGB stars \citep{kuchner02}, or stars polluted by the winds of a SAGB star and are now an AGB themselves \citep{maccarone16}, which may have similar nucleosynthetic signals to a \tzo.

There are a number of potential formation mechanisms of \tzos: engulfment of the NS during a common envelope \citep{taam78,terman95,ablimit22}; the NS receiving a kick such that it has a direct impact with the companion \citep{leonard94}; or a dynamical merger either in a dense stellar cluster \citep{ray87} or a triple system \citep{eisner22}. The formation of a \tzo{} is likely to produce a transient event \citep{hirai22}. Finally, \tzos{} are expected to die either when they run out of \texttt{rp}-seed material to burn forcing the material near the NS to heat up and enter the pair-instability region. This causes a runaway increase in the neutrino losses, causing the accretion onto the NS to no longer be Eddington limited, and the NS collapses into a BH \citep{Podsiadlowski95}. This may possibly form a transient event \citep{moriya18,moriya21}. Alternatively, they may eject their envelope via wind mass loss, leaving behind a ``bare'' NS \citep{bisnovatyi84}.

The number of \tzos{} in the Galaxy will depend on the rate of mergers, the fraction of mergers that successfully produce a \tzo{}, and the lifetime of the resulting \tzo{}. \citet{Podsiadlowski95} estimated a birth rate of $>10^{-4}\,\rm{yr}^{-1}$ from common-envelope evolution, and $\sim10^{-4}\,\rm{yr}^{-1}$ from NS kicks. \citet{renzo19} calculated the rate of collision from a NS being kicked during a SN with a companion to be $\approx 10^{-4}$ the core-collapse SN rate. While \citet{ablimit22} calculated formation rates of the merger of a ONeMg WD with a non-degenerate companion, where the core collapses and forms a NS, as between $\sim10^{-5}$--$10^{-4}\,\rm{yr}^{-1}$. Population synthesis calculations have shown an expected merger rate of a NS with a giant star as high as $\approx1$ per 100 core-collapse supernovae, depending on the assumed physics of the common envelope merger \citep{grichener23}. N-body simulations of globular clusters show dynamical mergers of a NS with main sequence star at a rate of the order 1 per $\sim4000$ NS containing binaries \citep{kremer20}. The actual formation rate of \tzos{} will however depend on the probability that the merger is successful and does not lead to a transient or complete disruption of the system \citep{schroder20}.

The first models of \tzos{} presented in \citet{thorne75,thorne77} were equilibrium models. Here the \tzo{} was split into three regions: an outer region which encompasses the envelope down to the knee; a middle region between the knee and the surface of the NS; and an inner region for the NS itself. These models were not evolved in time; instead, static solutions where found based on the assumed NS and envelope properties. \citet{biehle91} improved on the static models by including a simplified model of \texttt{rp}-burning, instead of just assuming CNO burning. Finally, \citetalias{cannon92} \& \citetalias{cannon93} created a set of evolutionary models where the NS was modelled by altering the EOS such that the electrons become increasingly degenerate in the vicinity of the NS. However, these models ignored wind mass loss and ignored changes in the composition.

In Section~\ref{sec:meth} we show how we build a \tzo{} in \MESA{} and discuss our default model. In Section~\ref{sec:evovle} we show the evolution of a grid of \tzo{} models. In Section~\ref{sec:pulse} we make predictions for the expected pulsation signal. In Section~\ref{sec:nuc} we explore the nucleosynthetic signal from \tzos{}. We discuss the suitability of our model assumptions in Section~\ref{sec:assume}, and show the possible final fates of \tzos{} in Section~\ref{sec:fate}. Finally, we discuss our results in Section  \ref{sec:discuss} and conclude in Section~\ref{sec:conc}.

\section{Building a \tzo{} model}\label{sec:meth}

Realistic modelling of the formation of a \tzo{} is a complex multi-dimensional problem involving the merger of a NS with the core of another star. In this work, we ignore both the NS and the merger process. Doing so, we construct a spherically symmetric post-merger structure of the envelope of the \tzo{} that is then allowed to evolve. There are three phases to creating our models: 1) producing an initial seed model, 2) modifying the inner boundary of our models, and 3) adjusting the global parameters of the model to match a chosen set of starting conditions. Our inlists and models are available at \url{https://doi.org/10.5281/zenodo.4534425}. An example model is also available in \MESA's \texttt{star/test\_suite} as of version a7c411b1.

\subsection{Seed model}

Using \MESA{} r22.11.1 \citep{paxton:11,paxton:13,paxton:15,paxton:18,paxton:19,jermyn22} we evolve a 20\msun{} star from the pre-main sequence until midway through the main sequence (MS). This model is used to initialise the stellar structure equations for all \tzos, independent of their final mass and composition. The precise choice of physics does not matter here, as the formation process will wipe away any knowledge the model had of its pre-\tzo{} state.

\subsection{Adjusting the inner boundary}

With our initial stellar model, we then model the NS by adjusting the inner boundaries of the stellar model. We gradually increase the inner mass, radius, and luminosity at the inner boundary of the model using \MESA's \texttt{relax\_core} method. This gradually changes the inner boundary while smoothly increasing the core density and core energy generation rate. By changing the inner mass boundary we do not change the total mass of the star, thus after forming a \tzo, we assume the total mass remains constant.

\subsection{Starting conditions}

With the base \tzo{} formed, we then further alter the model to match our required starting conditions. We use \MESA's \texttt{relax\_mass} to add (remove) mass via wind accretion (loss) to alter the total mass of the \tzo. We then change the initial metal composition \zinit, assuming a solar-scaled composition, and the initial helium fraction \yinit with \texttt{relax\_composition}. This allows us to build a \tzo{} with an arbitrary initial mass and composition, which can be used to approximate the post-merger structure for different formation scenarios.

We allow for arbitrary helium compositions, as the merger between a NS and a star may occur at any point in the star's lifetime (ignoring selection effects in terms of which stars are likely to successfully merge and form a \tzo). If a \tzo{} forms from a NS kick or a dynamical merger then the age of the companion star is a free parameter\footnote{Though this will be biased by the initial mass ratio and previous binary evolution.}, while mergers from a common envelope will likely, but not necessarily, involve the companion star having evolved beyond the MS. The mixing of the helium produced in the centre of a MS/post-MS star into the envelope will raise the helium mass fraction of the \tzo{} envelope compared to a canonical RSG. Mass loss from the envelope of the star during the merger will likely preferentially remove H-rich material and thus increase the average helium content of the \tzo{}. Thus when the star is fully mixed the average helium content over the entire star will increase. In the limit of when the entire hydrogen envelope is removed, a pure helium \tzo{} could be formed if it survives the merger process. However, whether a TZO may form from the merger of a NS with an evolved star or not is even more uncertain than \tzo{} formation with main sequence stars \citep{papish15,metzger22}.

At this point, we can also set other physics options, such as the choice of the nuclear network, accretion rate onto the NS, wind mass loss efficiency, or the mixing length. These are discussed further in Appendix~\ref{sec:others}. Although orbital angular momentum is involved in the merger process, the amount retained post-merger is still an open question \citep[e.g.,][]{schneider19}. Thus, for simplicity, in this work we consider only non-rotating models.

\subsection{Default \tzo\ model}

Here we describe our default \tzo{} model and the choices for uncertain physical and numerical quantities we have made. The effect of varying these choices is explored further in Appendix~\ref{sec:others}. We assume a default NS mass of $\mns=1.4\,\msun$. The radius of a NS is still uncertain and depends on the chosen NS equation of state (EOS) \citep{steiner10,miller19}. A realistic assumption would be to assume $R_{\rm{NS}}=10$--$20\,\km$ for the radius of the NS and use that as the inner boundary of our model\footnote{The inner boundary includes both the NS and the ``middle'' radiative burning region. However this radiative region is geometrically thin ($\sim100$m) and contains little mass ($\sim 10^{-8}\msun$) \citep{thorne77}.}. However, we have found that to be numerically difficult to model. Thus we move the inner boundary out to a radius of $\approx650\,\km$. This implies an average core density of $\avgrhoc=10^{9.3}\,\grampercc$, instead of $\avgrhoc\approx10^{14}\,\grampercc$ for a $10\,\km$ NS.

To compensate for the fact that we do not fully calculate the structure of the \tzo{} down to the surface of the NS, we inject energy at the inner boundary of our model. This energy injection approximates the missing energy generated from accretion and nuclear burning below the knee in the ``middle'' region. These assumptions are tested in Section~\ref{sec:assume}. We parameterise this inner energy injection based on the Eddington luminosity onto the NS as:

\begin{equation}
    \lknee = \eledd \ledd
\end{equation}

\noindent where the Eddington luminosity, \ledd, is:

\begin{equation}
    \ledd = \frac{4 \pi c G \mns }{\kappa_c}
\end{equation}

\noindent with $G$ the standard gravitational constant, $\mns$ is the mass of the neutron star, $c$ is the speed of light, and $\kappa_c$ of the opacity of the material at the inner boundary of the model. \lknee{} is allowed to evolve with time as the mass of the NS and the opacity at the inner boundary of the \tzo{} changes. Finally, $\eledd$ is an efficiency factor, for which by default we use $\eledd=1.0$. We use $L_{*}$ to denote the outer surface luminosity of our models.

By parameterising the energy in this way we can control how deep below the knee we model. The point where the local luminosity $L>\ledd$ is the point where radiation can no longer carry the energy and the envelope becomes convective. Thus when $\eledd=1.0$ the inner boundary of the structure we are calculating corresponds to the knee, which is also the base of the convection zone \citepalias{cannon93}. Smaller values of $\eledd$ include more of the material below the knee in the calculation but come at a greater computational cost. By injecting energy in this way we are assuming that the material below the knee is actually able to generate that much energy. It is possible that some of our models would not be able to do this, in which case the corresponding \tzo{} may not form. We are also assuming that the material below the knee is not mixed into the envelope, which may change the nucleosynthetic signatures.

We assume the NS grows at a rate set by the Eddington accretion rate on the NS:

\begin{equation}
  \medd = \emedd \frac{\ledd}{ c^2 }
\end{equation}

\noindent where \emedd is a scale factor which by default we set as $\emedd=1.0$. This leads to typical accretion rates of
$\sim 10^{-8}\,\msunyr$ for $\emedd=1$.

Our default model assumes an initial metallicity of $\zinit=10^{-4}$ and initial helium fraction $\yinit=0.28$. Our default \tzo{} has an initial total mass of $\minit=5\,\msun$, which includes a default NS $\mns=1.4\,\msun$, thus having an envelope mass of 3.6\msun. After 100 years of evolution, we enable \MESA's hydrodynamic capabilities \citep{paxton:15}. This wait is to allow the star to return to gravothermal equilibrium after the \tzo{} formation process. We also explore a series of model grids of \tzos{} with masses between $\minit=5$--20\msun, $\zinit=10^{-5}$--0.03, and $\yinit=0.28$--0.65.

We conservatively stop our models once the velocity of the surface layers exceeds 10\% of the escape velocity. At this stage the models undergo RSG pulsations, where there are large-amplitude surface pulsations which cause spiral patterns \citep{heger97} in a Hertzsprung-Russell diagram (HRD). This can change the stellar radius by factors of $\approx2$ over timescales of years. At this point the envelope begins expanding and contracting supersonically, and can reach $\sim40\%$ of the escape velocity before we can longer follow the evolution. These pulsations are resolvable in models of RSGs in \MESA{} when the timestep becomes much shorter ($\sim$days) than the pulsation timescale ($\sim$years) \citep{paxton:13}. The timestep becomes this small due to changes in the nuclear burning (see Section \ref{sec:oscil_rates}). While the evolution can be continued beyond this point by suppressing the hydrodynamics either globally or only in the cool outer envelope, the results become numerically unstable. More work is needed to determine how to couple the instabilities in the nuclear burning with the pulsational instabilities in the envelope.

We evolve the majority of our models using \MESA's \texttt{approx21.net} nuclear network. This contains the main PP, CNO, and alpha capture reactions up to \iron. This is a computational convenience. While \tzos{} are expected to undergo \texttt{irp}-burning, most of the energy is generated via CNO burning. We also test a large \nisos{} nuclear network for a limited set of models. This network was built by first adding all stable isotopes up to \ruthenium[]{}, then adding an additional neutron to the heaviest stable isotope for each element included. We then added all proton-rich isotopes with half-lives $>1$s. Finally, there was some hand tuning to make sure there were sufficient beta-decay pathways for all isotopes, as well as adding the light isotopes needed for PP, CNO, and the CNO breakout reactions. This network is available in the online Zenodo material.

We assume a mixing-length alpha parameter of $\mlta=1.8$, and use \MESA's time-dependent convection (TDC) mixing treatment \citep{jermyn22}. TDC has been shown to improve numerical stability during dynamical phases of evolution while reducing to the standard Cox-MLT prescription \citep{cox68} over longer timescales \citep{jermyn22}.
In models that have material below the base of the convection zone (when $\eledd < 1.0$), we assume that there is an additional weak mixing process occurring, with a diffusion coefficient of $10^6\, \rm{cm^2/s}$ in the material below the knee. This helps to prevent compositional gradients from building up near the surface of the NS \citep{piro07,keek09}. We do not include convective overshoot, semiconvective, or thermohaline mixing, as the star is almost fully convective.

Our EOS for the stellar component of the \tzo{} is a combination of \texttt{freeEOS} \citep{Irwin2004}, \texttt{HELM} \citep{Timmes2000}, and \texttt{Skye} \citep{jermyn21}. We make no assumptions about the NS EOS, as we can not make models with $10\lesssim R_{\rm NS}/\km \lesssim 20$, where NS radii are expected to be. We use the wind prescription of \citet{vanloon05}, which is based on observations of cool, dusty AGB \& RSG stars, with a wind scaling factor of $\etanl =1.0$.  This leads to typical wind mass loss rates of $\mdot\approx 10^{-5}$--$10^{-4}\,\msunyr$.

\MESA{} does not include any general relativistic (GR) correction factors. For this work, we add a GR correction factor to correct the continuity equation \citep{thorne77a,ayasli82}. This prescription adjusts the gravitational constant, $G$, as a function of the mass coordinate inside the star. All masses in this work are baryonic. Other \MESA{} choices are specified in Appendix~\ref{sec:mesa_other} and all options can be found in the online Zenodo inlists.

Our models have on average $\approx 1700$--$2000$ mesh points and we artificially cap the maximum timestep to be $\delta t= 2 \times 10^8\,\rm{s}$ ($\approx 6$ years). While models can take longer timesteps, this comes at the cost of an increased number of timesteps that needed to be rejected and taken again with a smaller $\delta t$ as \MESA{} could not find a valid solution that satisfies our required numerical constraints. Variations in spatial and temporal resolution of by factors of two smaller/larger lead to changes of order $\Delta \log \left(\rm{L_{*}/L_{\odot}}\right)\approx0.01\,\rm{dex}$.

\section{Structure and evolution of a \tzo{}}\label{sec:evovle}

\begin{figure}
    \centering
    \includegraphics[width=\linewidth]{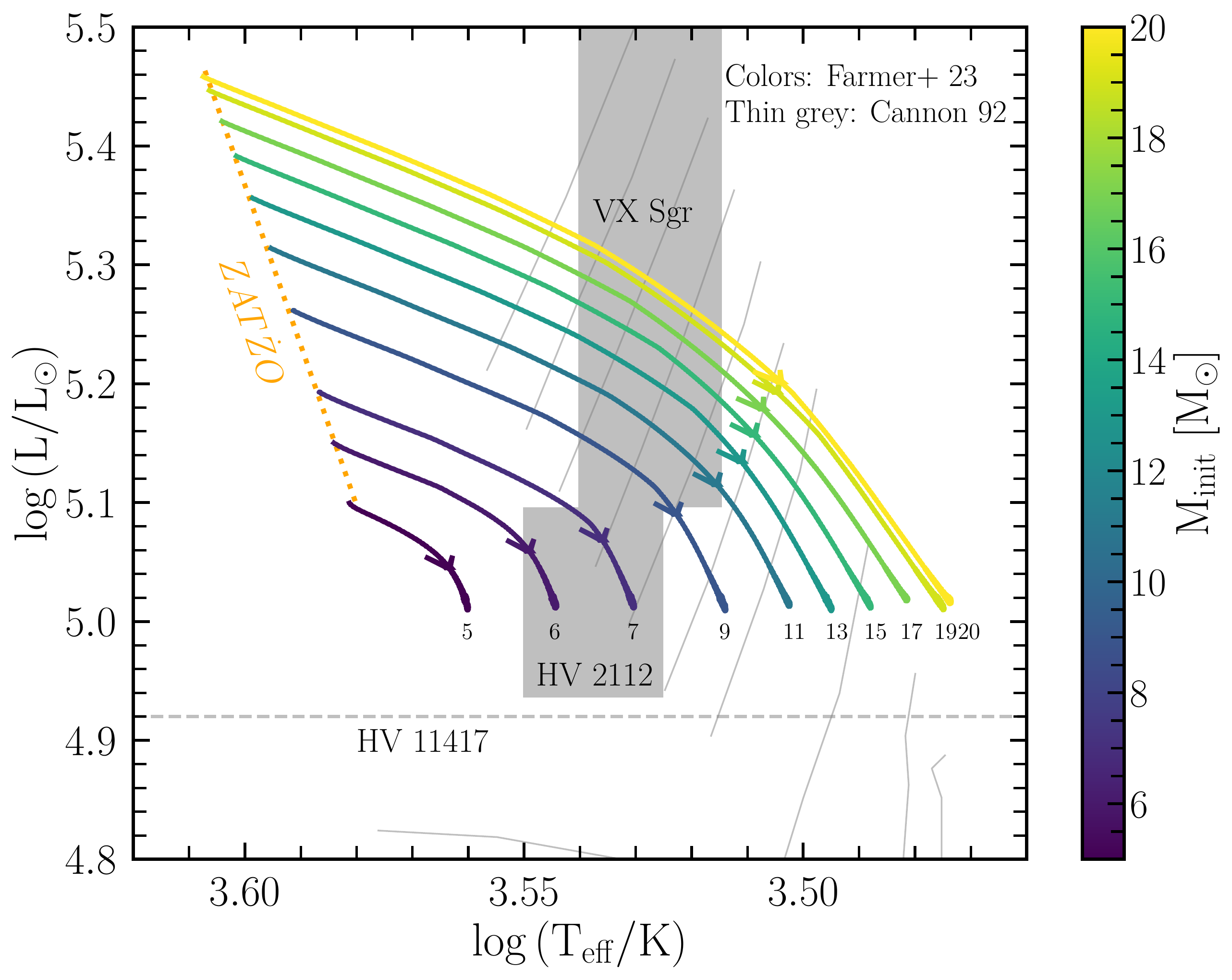}
    \caption{Hertzsprung-Russell diagram (HRD) of the \tzo{} as a function of their initial mass, at fixed initial composition. Models are evolved with our default assumptions and a $\mns=1.4\,\msun$. Colours indicate the initial mass of the \tzos. The direction of evolution is always towards \textit{decreasing} luminosity.
    The grey boxes show the observational constraints on HV 2112 \citep{levesque14} and VX Sgr \citep{tabernero21}. The horizontal dashed grey line shows the quoted luminosity of
    HV 11417 \citep{beasor18}, which has a variability amplitude of $\Delta m=1.86$ \citep{soszynski09}. The grey lines show the models of \citetalias{cannon92}, where evolution is always to \textit{increasing} luminosity. Arrows mark the midpoint of the \tzos{} lifetime.}
    \label{fig:hrd_int_mass}
\end{figure}

\begin{figure}
    \centering
    \includegraphics[width=\linewidth]{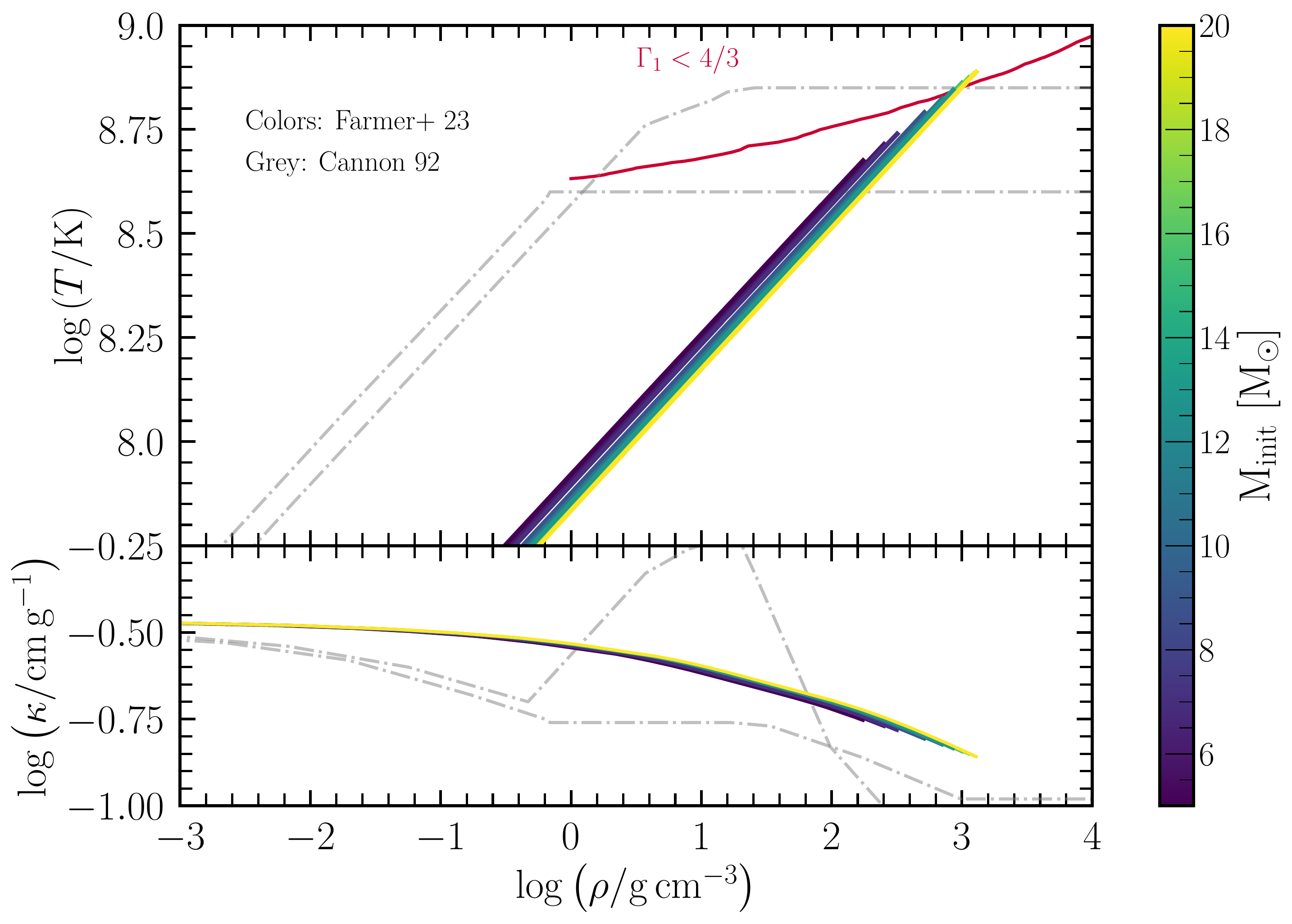}
    \caption{Top-panel: Temperature-density profiles our of our default \tzo{} models as a function of initial mass, at fixed initial composition. Models are evolved with our default assumptions and a $\mns=1.4\,\msun$. Colours indicate the initial mass of the \tzos. Dash-dotted lines show Model A and C from figures 6 and 7 of \citetalias{cannon92}.
    Bottom-panel: The opacity as a function of the density inside the star, dash-dotted lines are from tables 1 and 2 of \citetalias{cannon92}. Models shown are $\approx40,000$ years post \tzo{} formation. The red line marks the edge of the pair-instability region where electron-positron production becomes significant.}
    \label{fig:trho_int_mass}
\end{figure}

Figure~\ref{fig:hrd_int_mass} shows the evolution of the surface temperature and luminosity of our default \tzo{} models. Figure~\ref{fig:hrd_int_mass} shows the models once they are at least 100 years old (when we turn on the hydrodynamics), where they start from a zero-age \tzo{} (ZAT\.ZO) line. The endpoint of the models will be discussed further in Section~\ref{sec:fate}, as shown it is when the evolution reaches 1000 years before we can no longer follow the evolution. We can see that models with higher initial masses have higher initial luminosities and temperatures compared to lower-mass models. Higher-mass models also evolve to lower final surface temperatures and live longer.

Contrary to the models of \citetalias{cannon92} our models always evolve to lower luminosities. The total luminosity of our \tzos{} is a combination of the injected energy, \lknee{}, and nuclear burning above the knee. The nuclear burning above the knee provides $\approx 1\%$--$10\%$ of the total luminosity, and this fraction decreases with time as the knee cools. Thus the evolution is driven by changes in \lknee, where $\lknee \propto \mns/\kappa_c$. The accretion rate on to the NS is $10^{-9}$--$10^{-8}\, \msunyr$, which combined with a lifetime of $\approx100,000$ years means the NS can only gain $10^{-4}$--$10^{-3}\,\msun$. Thus the evolution of our \tzos{} can be approximated as entirely driven by the increase in the opacity of material at the knee, which lowers \lknee{}.

The opacity for the material at the knee is provided by Compton scattering \citep{poutanen17} and thus depends on the number of free electrons per nucleon. Three factors will drive the evolution of the number of free electrons: H burning into He will decrease the number of free electrons, proton captures in \texttt{rp}-burning will decrease the number of nucleons, and $\pair$ production will increase the number of free electrons. If we turn off composition changes from nuclear burning (by setting \texttt{dxdt\_nuc\_factor}=0) while preserving the energy generated from nuclear burning, we find \lknee{} still decreases. However, it does this at a much faster rate than when the composition is allowed to change. Therefore, nuclear burning slows the decrease in \lknee{} but can not stop it. Thus, the changes in \lknee{} are driven by the production of \pair{} pairs, while not entering the pair-instability region.

Our results differ from those of \citetalias{cannon92} due to differences in the chosen stopping criteria. Our models stop when supersonic pulsational instabilities form in the envelope, preventing \MESA{} from continuing the evolution. The end condition for \citetalias{cannon92} is when the NS grows to the Oppenheimer-Volkoff (OV) \citep{oppenheimer39} mass limit for the their assumed EOS, which for \citetalias{cannon92} is 2\msun. As the NS mass increases, \lknee{} will increase, increasing the total luminosity of the star in agreement with the results of \citetalias{cannon92}. However, our models stop significantly earlier than \citetalias{cannon92} such that the NS only gains $10^{-4}$--$10^{-3}\,\msun$, which does not cause \lknee{} to increase by a significant amount. We note in passing that \citetalias{cannon92} states they can take single timesteps of $\approx10^5$ years, which is comparable to the entire lifetime of many of our models which is between 50,000 and 200,000 years for the \tzos{} shown in figure \ref{fig:hrd_int_mass}.

We can replicate the results of \citetalias{cannon92} by increasing the accretion rate onto the NS. If the accretion rate onto the NS is high enough that the mass of the NS can grow significantly (at least $\approx0.1\,\msun$) then \ledd{} will increase in time. This causes the \tzo{} to become more luminous and more closely follow the tracks of \citetalias{cannon92}, though still at higher surface temperatures and lower surface luminosities. This requires accretion rates of between $10^3$--$10^4\,\medd$ (See Appendix~\ref{sec:others}). The remaining differences between our models can likely be attributed to changes in the microphysics (EOS, opacities, and nuclear reaction rates) and choice of metallicity.

Figure~\ref{fig:trho_int_mass} shows the temperature-density profile and opacity-density profile inside our \tzos{} arbitrarily at $\approx40,0000$ years after \tzo{} formation. As the initial mass of the model increases the density also increases and the models reach higher temperatures at the base of the convection zone. Our models do not form a knee, due to the energy injection at the inner boundary of the models, so we are only modelling material above the knee. The dash-dotted lines show models A \& C of \citetalias{cannon92}. We can see that our models are significantly denser (by a factor $\approx100$) than previously predicted, though this is still a factor $\approx100$ less dense than a typical AGB/RSG.

None of our models ever evolve significantly into the pair-instability region (where $\Gamma_1 < 4/3$). Instead, models evolve \textit{around} the edge of the pair-instability region (See Section~\ref{sec:assume}). Note that the temperature and density of the knee evolves with time, but always stays outside of the pair-instability region.

The bottom panel of Figure~\ref{fig:trho_int_mass} shows the opacity of our models in colour, and models A \& C of \citetalias{cannon92} with grey dash-dotted lines. When $\log (\rho/\grampercc) \leq 0.0$ our models have slightly higher opacities, as they are cooler for the same density. Once $\log (\rho/\grampercc)>0.0$ our models have higher opacities than model A, but lower opacities than those of model C (which enters the pair-instability region and so \pair{} production dominates the opacity).

In \citetalias{cannon93} they find a set of ``high'' and ``low'' mass solutions, which depends on how the energy is generated at the knee, and the production of \pair{} due to models entering the pair-instability region which changes the required \lknee{} as the opacity increases (these are also the ``giant'' and ``supergiant'' models of \citealt{thorne77}). \citetalias{cannon93} finds a gap where, for certain masses, models are unable to produce sufficient energy to keep the envelope convective. Our models do not exhibit this ``luminosity gap'' (which can be mapped into a mass gap) in the mass distribution of \tzos{}. This is due to the higher density of our models, such that they do not enter the pair-instability region. Therefore our models would not be limited by the same mechanism as proposed by \citetalias{cannon93} and thus we should not expect a split into ``high'' and ``low'' mass solutions, even if we did model the knee. \citetalias{cannon92} also finds valid solutions for all masses, where \citetalias{cannon93} attributes this difference due to changes in the nuclear reaction rates used.

\subsection{Composition effects}\label{sec:comp}

\begin{figure}
    \centering
    \includegraphics[width=\linewidth]{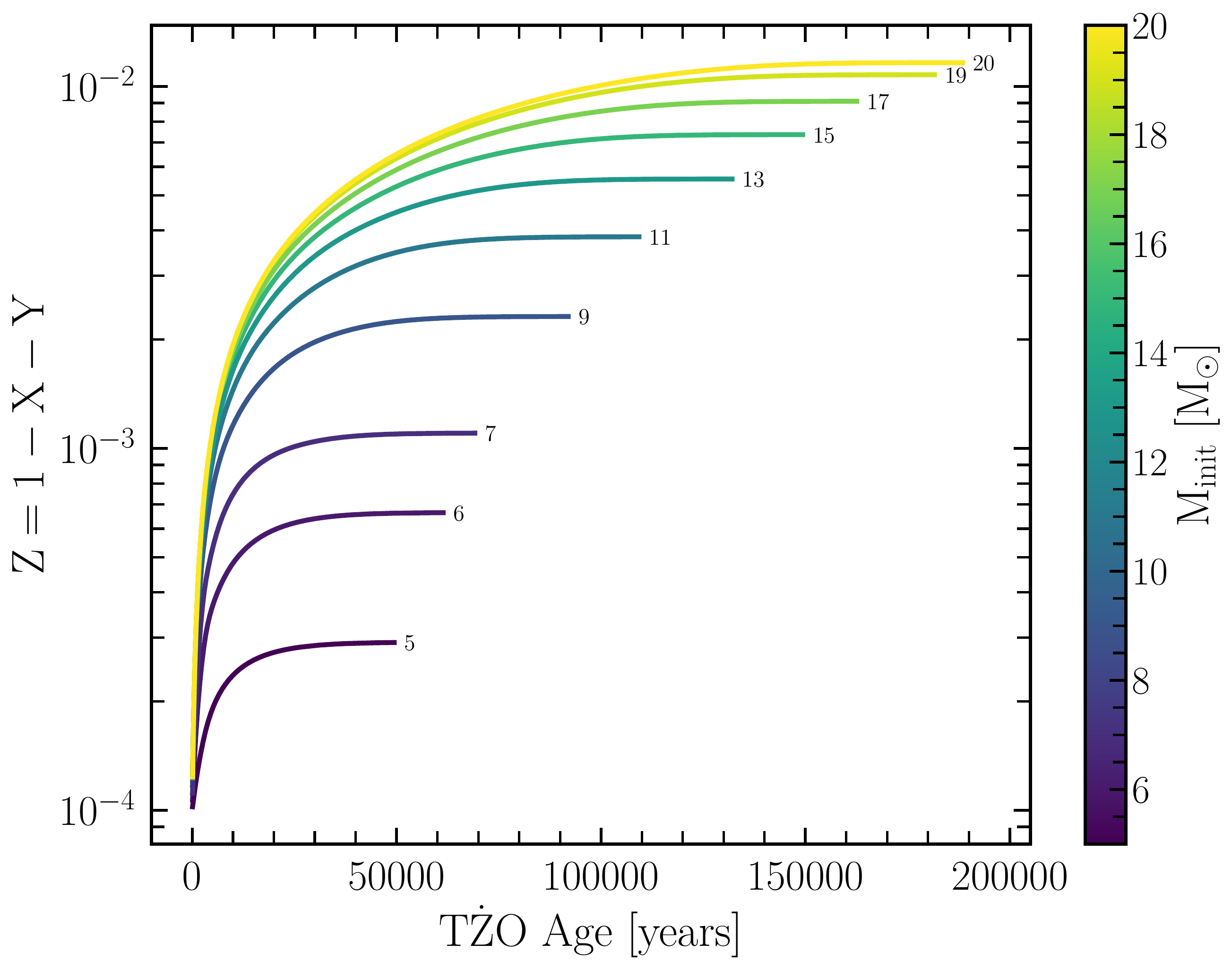}
    \caption{The evolution with time of the average metal fraction (Z) as a function of the initial mass, at fixed initial composition. Models are evolved with our default assumptions and $\mns=1.4\,\msun$. Colours indicate the initial mass of the \tzos. The final ages of these models are only lower limits to the lifetime of a \tzo{}.}
    \label{fig:avgZ_int_mass}
\end{figure}

Figure~\ref{fig:avgZ_int_mass} shows the average metal fraction of our \tzos{} as a function of time since the \tzo{} formed. These models start at $Z=10^{-4}$ but can increase their total metal fractions by factors of 10-100 within $\approx10,000$--100,000 years. Higher initial mass models evolve to higher total metal fractions, reaching near solar values for their total metallicity, though this is a significantly non-solar scaled composition. They also evolve faster to higher metallicities than the low-mass models and live longer. The ages shown are only lower limits on the lifetime of a \tzo{} (see Section~\ref{sec:fate}), but likely lead to lifetimes of $\approx 10^4$--$10^5$ years.

\begin{figure*}
     \centering
     \subfigure[Initial metallicity]{\label{fig:hr_z}\includegraphics[width=0.49\linewidth]{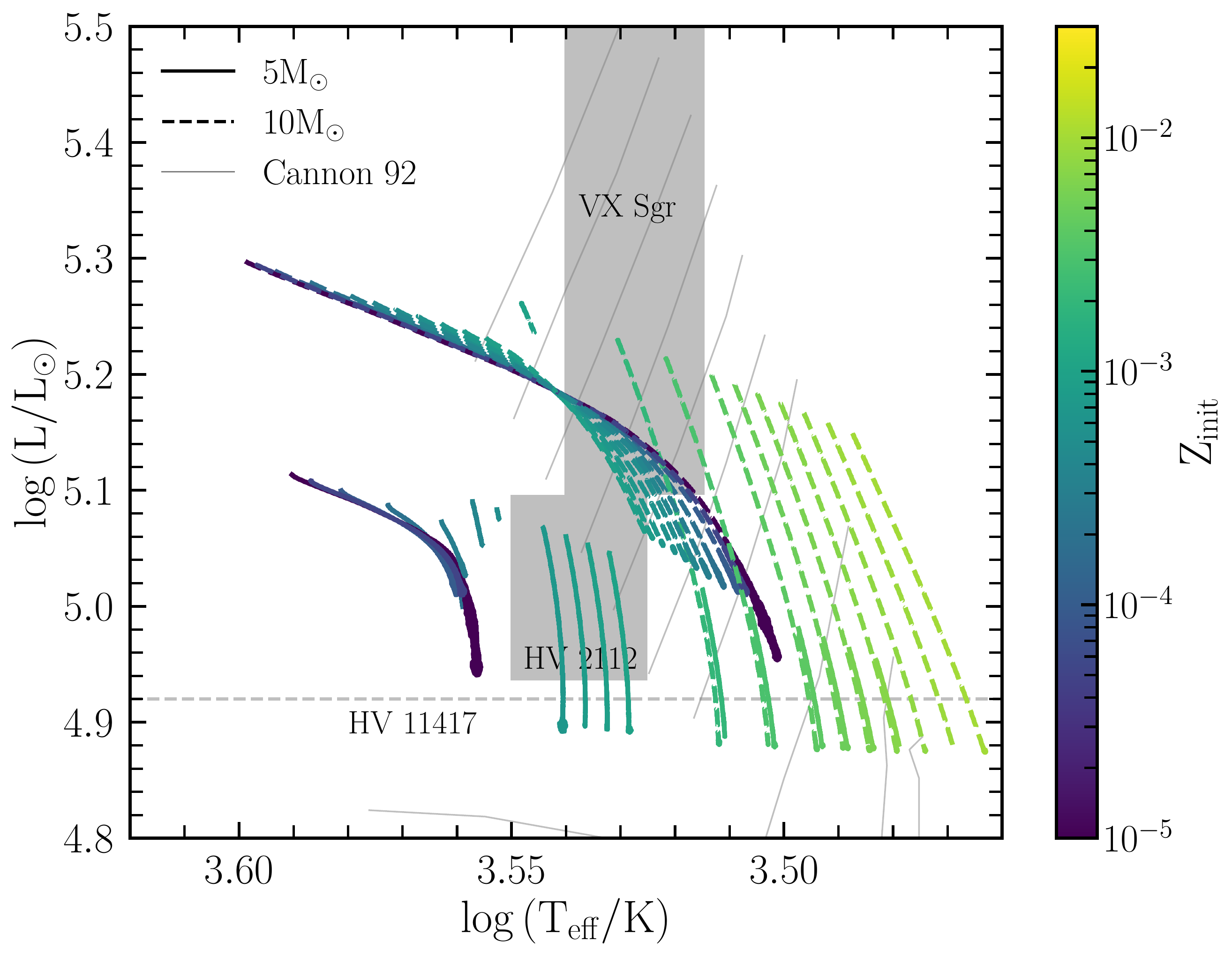}}
     \subfigure[Initial helium fraction]{\label{fig:hr_y}\includegraphics[width=0.49\linewidth]{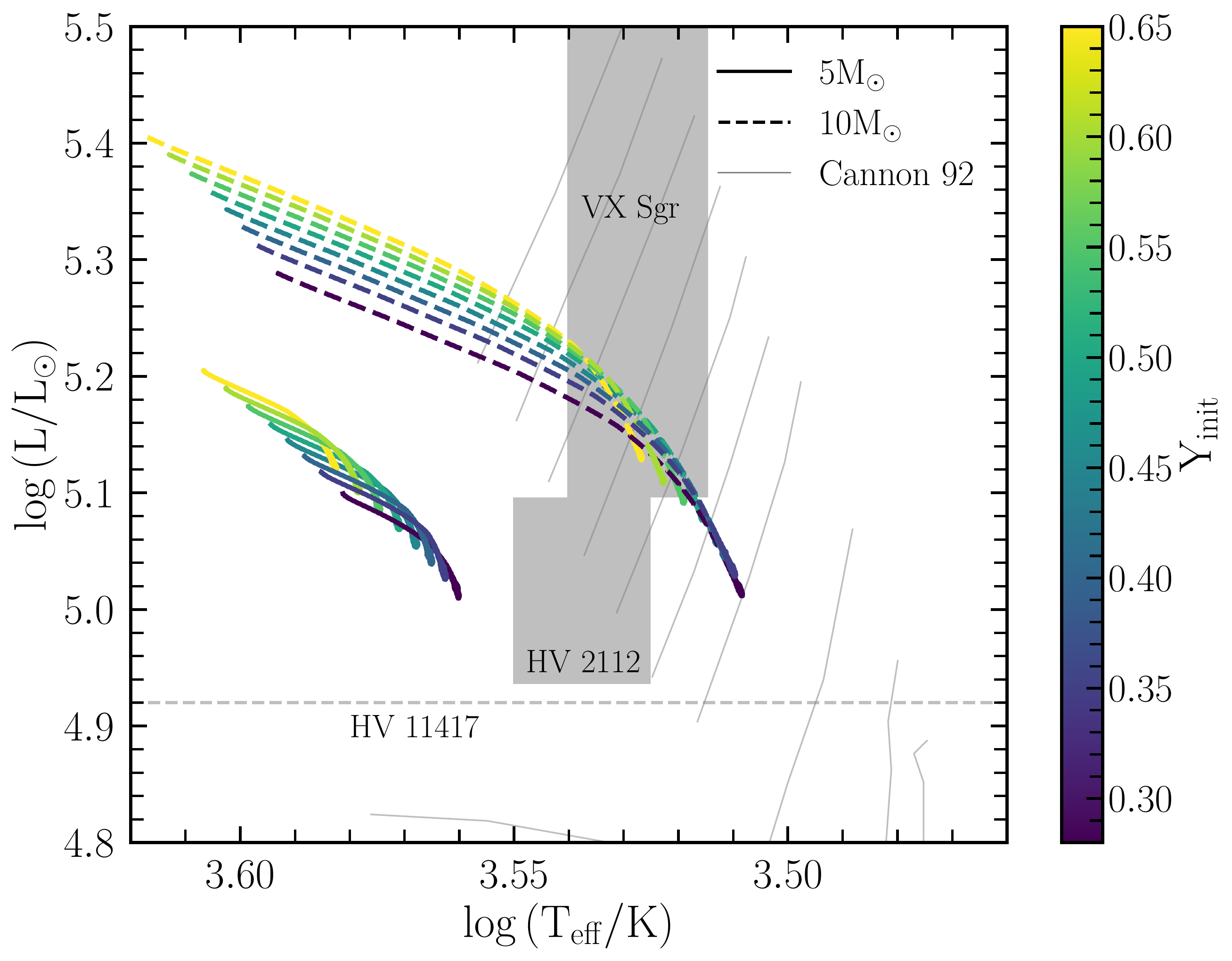}}

        \caption{HRD for variations in the initial metallicity (panel a) and initial helium fraction (panel b). Solid lines denote a $5\,\msun{}$ \tzo{}, while dashed lines denote a $10\,\msun{}$ \tzo.
        Grey lines and grey boxes have the same meaning as in Figure~\ref{fig:hrd_int_mass}.}
        \label{fig:hr_comp}
\end{figure*}

Figure~\ref{fig:hr_z} shows the HRD of \tzos{} with varying $\zinit$ with initial masses of 5 and 10\msun{}. The \tzos{} can be broken into two groups of behaviour, a low initial metallicity and a high initial metallicity behaviour. The exact initial metallicity at which this behaviour changes depends on the initial mass, for the 5\msun{} models this is at $\zinit \approx 3\times10^{-4}$ while for the 10\msun{} models this occurs at $\zinit\approx 10^{-3}$. This limit is approximately the final metallicity shown for each mass in Figure~\ref{fig:avgZ_int_mass} (though the match is not exact). Effectively the low-metallicity models evolve up to the mass dependent critical $Z$, with the evolution converging to the same final surface temperature and luminosity once they reach the critical $Z$. In contrast, the models that start with metallicities above this critical limit follow their own $Z$-dependent evolutionary tracks. We can also see in Figure \ref{fig:hr_z} that the high-metallicity models evolve along similar tracks independent of the initial mass (except for their starting luminosities). We also note that as the initial metallicity increases the 5\msun{} models are more likely to become numerically unstable.

As the metallicity increase (either due to a higher initial metallicity or due to metals produced from nuclear burning), the opacity at the knee increases \citep{xin22}. This decreases \lknee{} and thus decreases the temperature at the knee. This decreases the amount of heavy metal burning as the metallicity increases, though the rate of CNO burning increases as the metallicity increases due to the increased amount of CNO material. Thus the nuclear burning luminosity is greater in the high metallicity models, but the rate of production of all metals is lower. \textit{Hence at LMC and SMC metallicites \tzos{} may show little metal enrichment.} The high metallicity models can have lifetimes up to a factor $\approx2$ more than shown in Figure~\ref{fig:avgZ_int_mass}.

Figure~\ref{fig:hr_y} shows variations in the initial helium fraction \yinit, for a 5 and 10\msun{} \tzo{} at $\zinit=10^{-4}$. As the initial helium fraction increases, the models become more luminous and have higher surface temperatures. The variation in the initial helium fraction may be expected due to variations in the evolutionary state of the companion when the \tzo{} merger takes place. As \yinit{} increases the final luminosity they reach increases and is at higher surface temperatures. These changes are due to the knee decreasing in temperature as \yinit{} increases. This temperature decrease also decreases the rate of metal production.

The upper limit of $\yinit=0.65$ in Figure~\ref{fig:hr_y} is due to numerical issues, which prevents the modelling of \tzos{} with $\yinit=0.65$--0.95.
While there are differences between H-rich and He-rich \tzos{}, they both occupy similar regions of a HRD, appearing as RSGs. It is interesting to speculate that, if more He-rich \tzos{} can form, there may exist a population of H-poor RSGs. While we are unaware of any He-lines detectable in a cool RSG atmosphere, H-lines are readily detectable. Thus weak or no detectable H-lines in a RSG spectrum could indicate a very He-enriched \tzo{}.

\subsection{Comparison with proposed \tzo{} candidates}

Here we briefly compare observed \tzo{} candidates with our models at both our default composition (Fig.~\ref{fig:hrd_int_mass}) and for alternative compositions (Fig.~\ref{fig:hr_comp}). We emphasize that our default model has $\zinit=10^{-4}$, i.e.\ much lower than that of the SMC or LMC.

HV 2112's temperature and luminosity can be well approximated by models using with an initial mass between $ 5 \lesssim \mint/\msun{}\lesssim 8$ when adopting our default composition. HV 2112 can also be well fitted by our 5\msun{} \tzo{} models with $\zinit\approx10^{-3}$, though for models at this metallicity we would not expect to see any metal enrichment (see Section \ref{sec:comp}). Thus while HV 2112 was initially proposed as a \tzo{} candidate due to its unusual chemical composition, \emph{we believe this unusual composition now rules out HV 2112 being a \tzo}.

VX Sgr's position in the HRD can be matched with a \tzo{} model of mass $\minit \gtrsim 8\,\msun$ at our default metallicity, and is also consistent with our $10\,\msun$ models with modified composition.

HV 11417 has a luminosity which is lower than any of our default-composition models in Fig.~\ref{fig:hrd_int_mass}. However, lower-mass NSs can allow models to drop to the luminosity of HV 11417 with our default composition (see Appendix~\ref{sec:others}), and several of the higher-metallicity $5\,\msun$ models shown in Fig.~\ref{fig:hr_comp} reach the luminosity of HV 11417 even for our canonical NS mass.

\section{Pulsations}\label{sec:pulse}

\begin{figure}
    \centering

    \subfigure[]{\label{fig:pulse1}\includegraphics[height=0.23\paperheight]{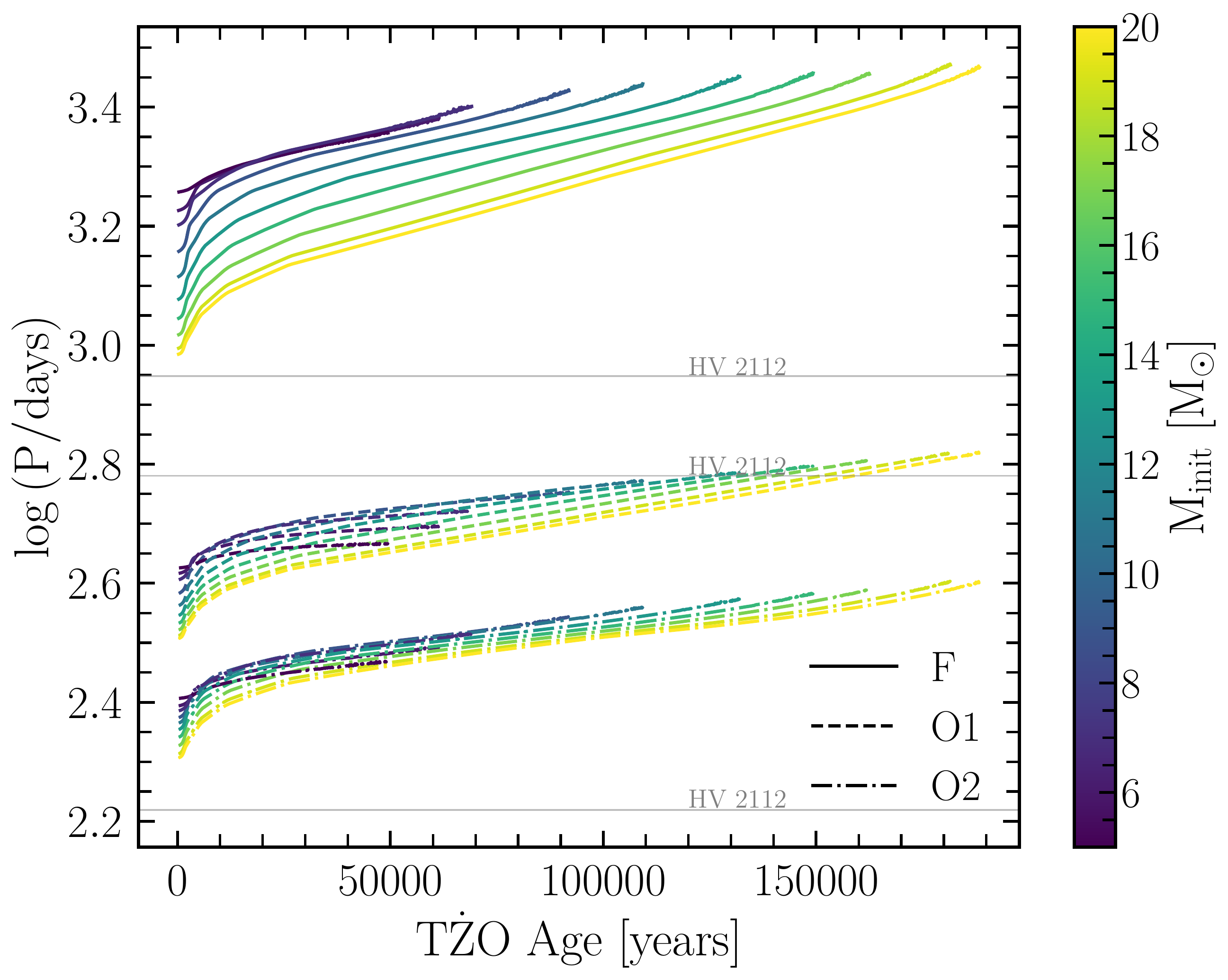}}
    \\
    \subfigure[]{\label{fig:pulse2}\includegraphics[height=0.23\paperheight]{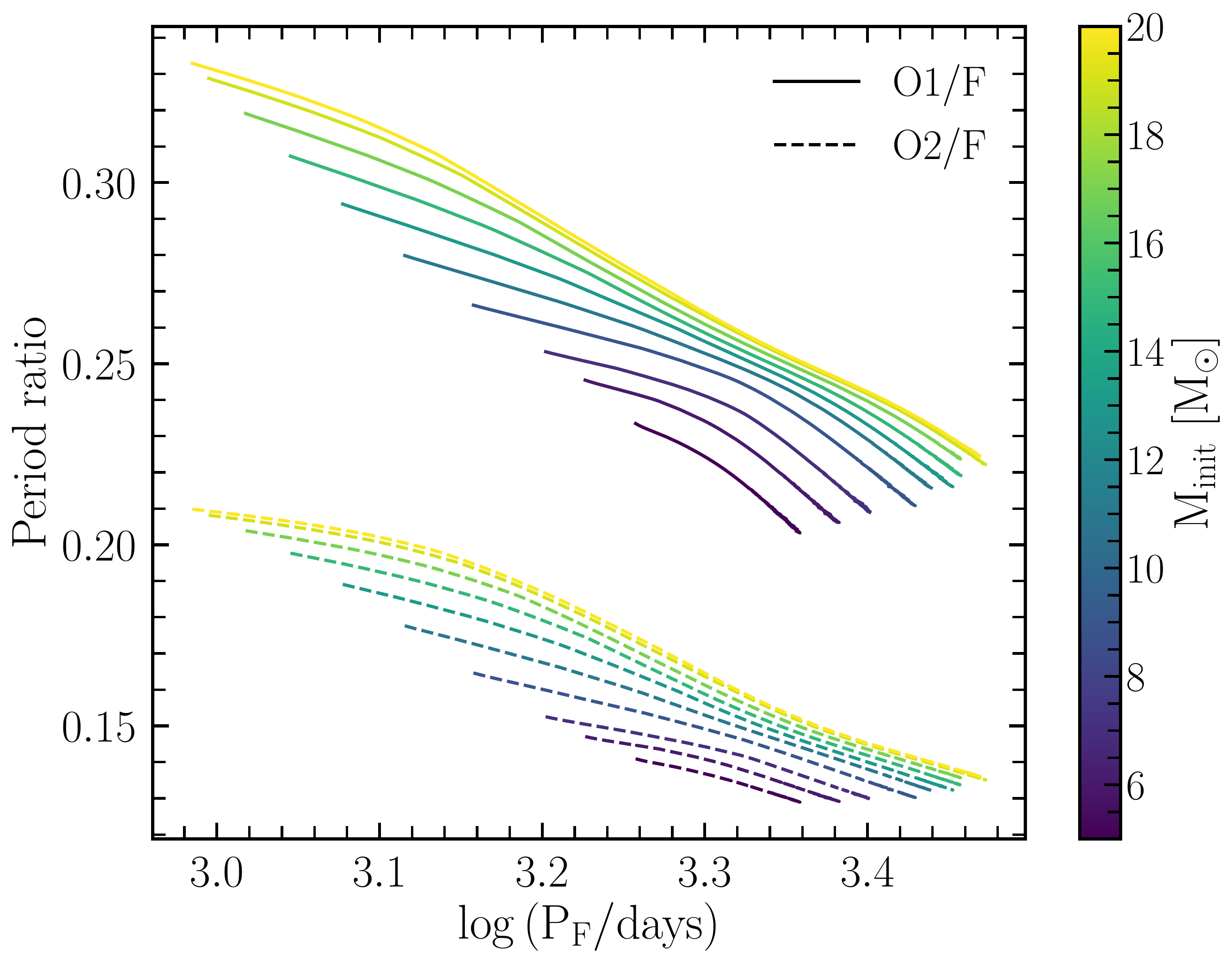}}
    \\
    \subfigure[]{\label{fig:pulse3}\includegraphics[height=0.23\paperheight]{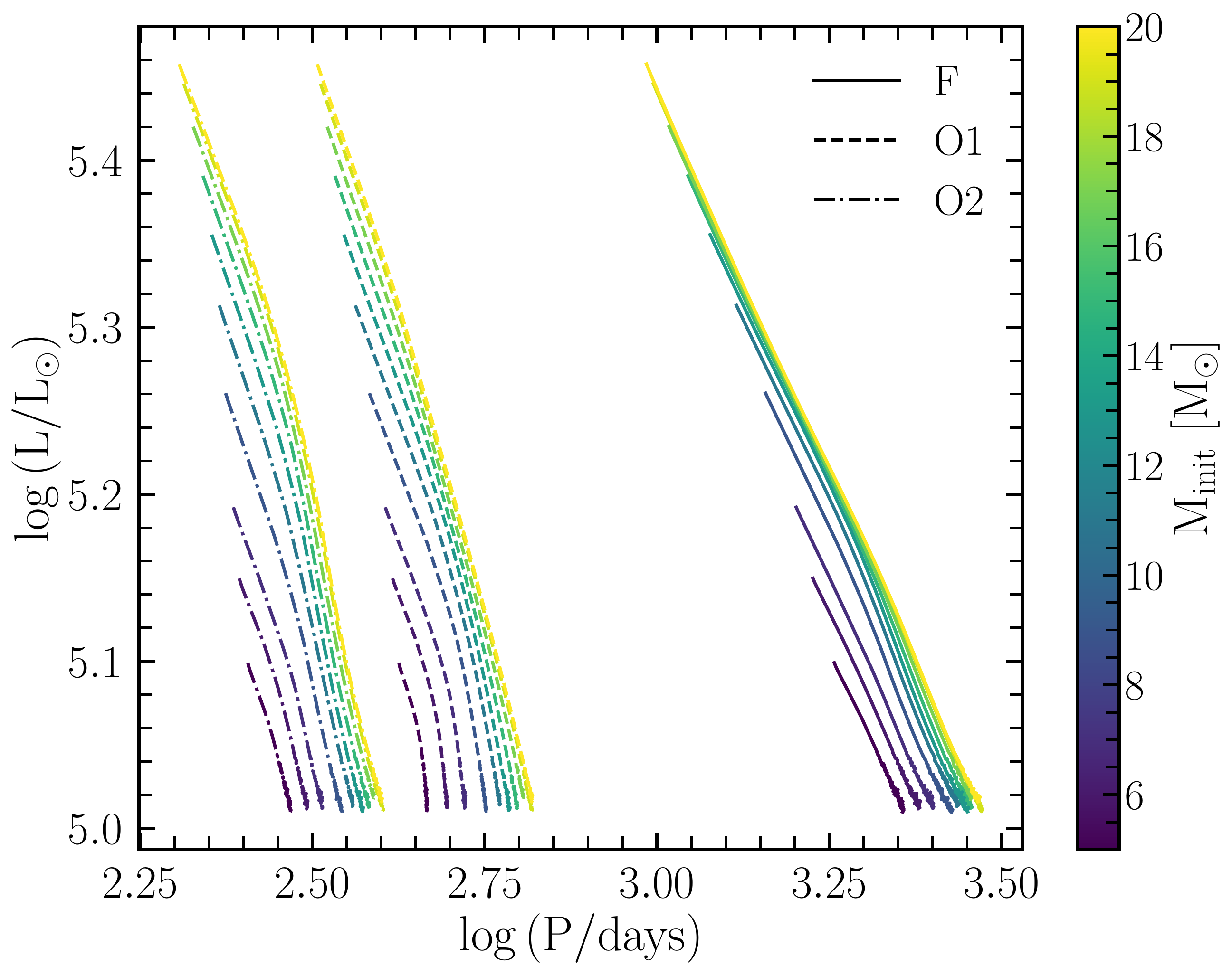}}
    \caption{Pulsation properties of our \tzos{} with default model assumptions.
    Panel a: the evolution of the pulsation period as a function of time for the fundamental (solid), first (dashed), and second overtones (dot-dashed). Panel b: Petersen diagram showing the ratio of the first overtone to the fundamental (solid), and the second to the fundamental (dashed). Panel c: The pulsation period as a function of the surface luminosity, lines have the same meaning as panel a. Colours indicate the initial mass of the \tzo.
    Grey dashed lines in panel a mark the observed pulsation periods of HV 2112 \citep{soszynski11}.}
    \label{fig:pulse_int_mass}
\end{figure}

\subsection{Hydrostatic pulsations}\label{sec:pulse_static}

RSGs may pulsate due to the $\kappa$-mechanism in hydrogen ionisation zones, with periods between $\approx 100$--1000 days \citep{fox82,heger97}. Longer period pulsations may also occur, but observations can be hindered by a lack of long baseline photometry \citep{soraisam18}.
Observations of RSG pulsations are complicated by irregular photometric variability, presumed to be caused by the interaction of large convection cells (which can have sizes of order the stellar radius) and the pulsation modes \citep{kiss06}.

Using the \GYRE{} v6.0 \citep{Townsend13, Townsend18} stellar oscillation code, we compute the adiabatic pulsations for our \tzos{} with $\minit=5$--$20\,\msun$ for the radial ($l=0$) modes. Figure~\ref{fig:pulse1} shows the time evolution of the pulsation period, Figure~\ref{fig:pulse2} shows a Petersen diagram \citep{petersen73} of the period ratios, while Figure~\ref{fig:pulse3} shows the period-luminosity relationship.

In \ref{fig:pulse1} we can see that the fundamental mode is always greater than 1000 days, and this increases with time. The first and second overtones are $\approx500$ and $250$ days respectively. The highest-mass models start with lower periods than the low-mass models. While exploring other parameters, we find that the period of the fundamental mode is predominately sensitive to \yinit{}, potentially providing a way to constrain the helium composition of a \tzo.

Figure~\ref{fig:pulse2} shows a Petersen diagram of the period ratios compared with the fundamental period. We can see that in general the period ratio is small ($\approx0.2$--0.3), decreases as the period increases (and thus decreases with time), as well as decreasing with initial mass. OGLE observations of the SMC and LMC \citep{soszynski04}, show most objects with $\log (\rm{P/days})>3$, have period ratios $\approx0.1$. Therefore our \tzos{} exist in a region of parameter space distinct from most other objects, potentially providing an easy method to search for \tzo{} candidates.

Finally, Figure~\ref{fig:pulse3} shows the period-luminosity relationship for our \tzos{}. We can see a tight relationship between their pulsation period and luminosity. Thus \tzos{} may prove useful for determining distances if they can be detected and identified.

HV 2112 has a measured set of pulsation periods from OGLE of 165.59, 604.4, and 887.3 days \citep{soszynski11}. From Figure~\ref{fig:pulse1} it is difficult to match the models to the measured periods of HV 2112. To match the 604.4 and 887.3 day periods will require the first and second overtones to increase their periods by a factor of $\approx 2$.

Variations in other model parameters (Appendix~\ref{sec:others}) lead to changes in the predicted pulsation periods. However, no parameter variation we have tested is able to fit the measured periods of HV 2112, over the time we have evolved our \tzos{} for (though our grid is not exhaustive). The pulsation periods are predominately set by the initial mass, the mass of the NS, \mlta, and the initial helium fraction, while not being sensitive to the initial metallicity. The fundamental mode is most strongly affected by \yinit{}, with \yinit=0.65 able to have $P>10^5$ days, while the first overtone is more strongly affected by \mlta.

The closest match in our (non-comprehensive) grid is from the 5\msun{} \tzo{} with $\mlta=3.0$, with a first overtone period of $\approx750$ days. Based on 3D models values of $\mlta=3$--4 have been proposed for RSG envelopes \citep{goldberg22}. We speculate a slightly more massive \tzo{} with a relatively high \mlta{} would be able to better fit the pulsation periods of HV 2112. If HV 2112 were a \tzo{} with a high \mlta, then we would predict that we have observed the first, second, and third overtones, and that the fundamental mode is currently undetected with a period in the 1500--3000 day range.

VX Sgr has a measured pulsation period of 757 days, with a possible longer period at 28,279 days \citep{tabernero21}. Though \citet{tabernero21} caution the 28,279 days is comparable to the total time span that VX Sgr has been observed for, and thus may be an artifact of the observing period. There are also a number of shorter pulsations that vary in period and amplitude, around $~600$ days. The lack of a detected period between 1000-3000 days rules out most of our \tzo{} models. The only models that can match the 28,279 day period (assuming that it is real),
is our most helium enriched model with $\yinit=0.65$. This model has a fundamental mode that increase with age and can be  between $P\approx10^3$--$10^5$ days. Thus if VX Sgr where a \tzo{} it would imply the merger occurred with an evolved star after significant amounts of helium had been produced in the companion.

HV 11417 has measured pulsation periods of 214, 793, and 1092
  days \citep{soszynski09,soszynski11}. Based on Figure
  \ref{fig:pulse_int_mass}, the 1092 day period would imply it was a
  very young ($\lesssim2000$ year) and massive ($\sim20\msun$) \tzo{}.
  However models in this mass and age range would be inconsistent with
  the reported luminosity of HV 11417 by $\sim0.6$ dex. Thus we
  conclude HV 11417 is inconsistent with our a \tzo{} models.

\subsection{Oscillations in the nuclear burning rate}\label{sec:oscil_rates}

\begin{figure}
    \centering
    \includegraphics[width=\linewidth]{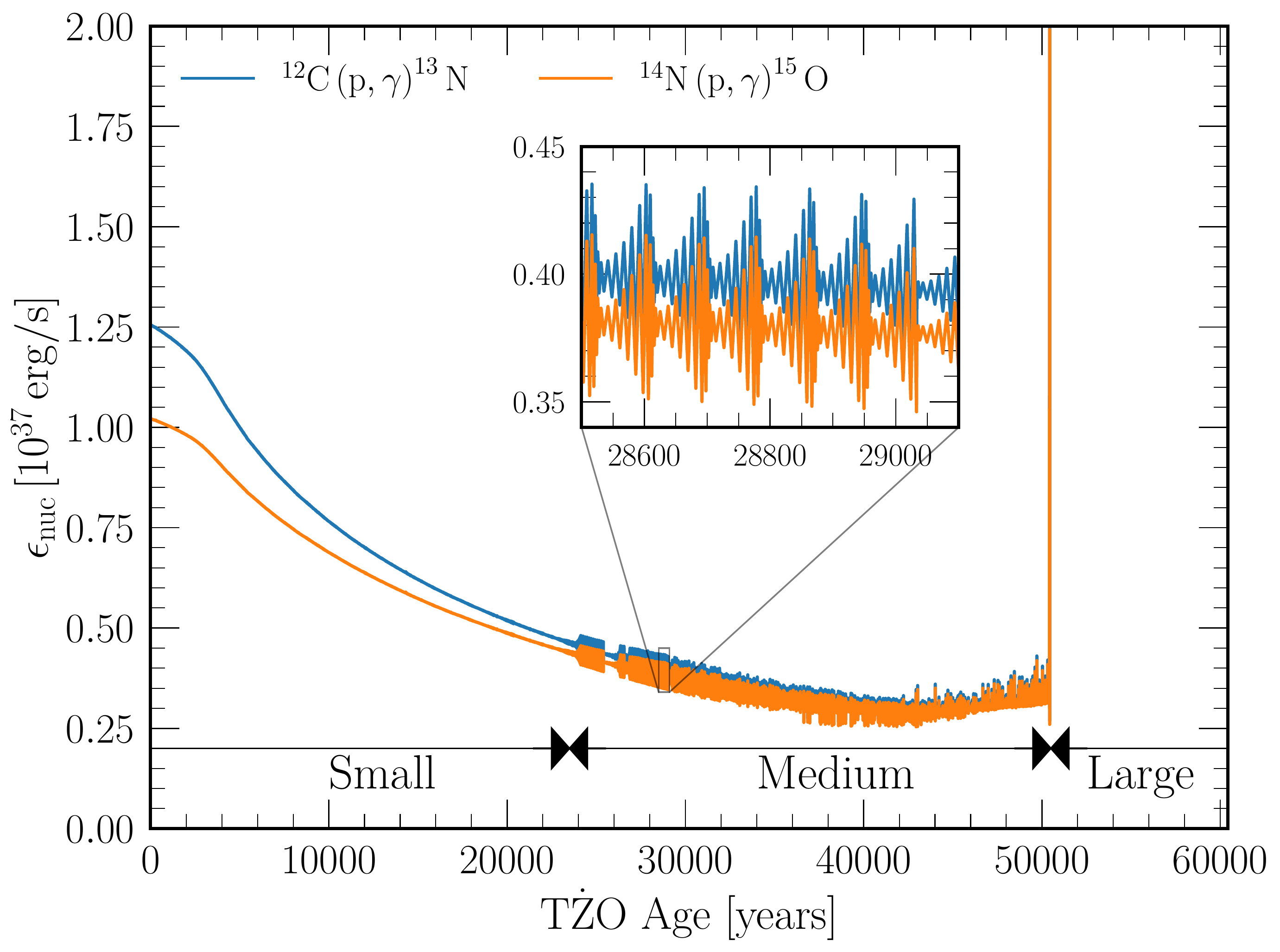}
    \caption{The nuclear energy released, integrated over the \tzo{} for the two most energetic reactions in our default 5\msun{} \tzo{} evolved with the \texttt{approx21.net}. These two reactions provide $\approx85\%$ of the star's nuclear energy. The inset shows a zoom-in where the reaction rates begin to undergo medium oscillations. These oscillations are still within the \tzosp{} hydrostatic phase of evolution. At $\approx 50,000$ years the nuclear energy generation rate increases rapidly and exceeds the plot limits while becoming hydrodynamic.
    Text labels denote the approximate boundaries between the different amplitudes of the pulsations.}
    \label{fig:pulse_rates}
\end{figure}

Figure~\ref{fig:pulse_rates} shows the total nuclear energy integrated over our default 5\msun{} \tzo{} for the $\rm{\carbon[12](p,\gamma)\nitrogen[13]}$ and $\rm{\nitrogen[14](p,\gamma)\oxygen[15]}$ reactions. These two rates constitute ~85\% of the nuclear energy generated in our models (with the \texttt{approx21.net}) network. As the \texttt{approx21.net} network approximates CNO burning, $\rm{\carbon[12](p,\gamma)\nitrogen[13]}$ is a proxy for the chain $\rm{\carbon[12](p,\gamma)\nitrogen[13](,e^{+}\nu)\carbon[13](p,\gamma)\nitrogen[14]}$, while $\rm{\nitrogen[14](p,\gamma)\oxygen[15]}$ proxies the chain $\rm{\nitrogen[14](p,\gamma)\oxygen[15](,e^{+}\nu)\nitrogen[15]}$.

When the \tzos{} are $\lesssim 24,000$ years old the variations in the reaction rates are small and approximately the size of the linewidth in Figure~\ref{fig:pulse_rates}. Once the \tzo{} reaches $\approx 24,000$ years the reaction rates begin showing a medium level variability, with a short period variability of a few years embedded in a longer $\approx 100$ year period cycle (which is shown in the insert). This leads to a variation in the surface luminosity of $\Delta \logl \approx 0.001$ dex. Then when the \tzo{} reaches $\approx 50,000$ years, the model undergoes large variations in the reaction rates. The energy generation rate increases by a factor $\approx15$ to $\approx10^{38}\,\rm{erg/s}$ and the surface luminosity can vary by $\Delta \logl \approx 0.5$ dex. This is when we can no longer follow the evolution with  hydrodynamics included in the model,

We can locally suppress the hydrodynamics in the envelope by using \texttt{velocity\_logt\_lower\_bound=5}, which turns off the hydrodynamics in zones where the local temperature is $\logt < 5.0$ which allows the evolution to continue for another $\approx40,000$ years. As the \tzo{} evolves at this point the nuclear energy generation rates drops, returning to its pre-large amplitude burning values, before once again becoming numerically unstable and the evolution ceases.

\subsection{Hydrodynamic pulsations}\label{sec:pulse_dyn}

Convective envelopes are known to be able to have dynamical instabilities that can lead to mass loss \citep{pacynski69,tuchman78,tuchman79}. Given a suitable excitation these dynamical instabilities can lead to mass loss rates of $\sim 10^{-3}\msunyr$ through pulsational mass loss events \citep{clayton17}. Here we show what happens to our \tzos{} as they undergo RSG pulsations.

\begin{figure}
    \centering
    \includegraphics[width=\linewidth]{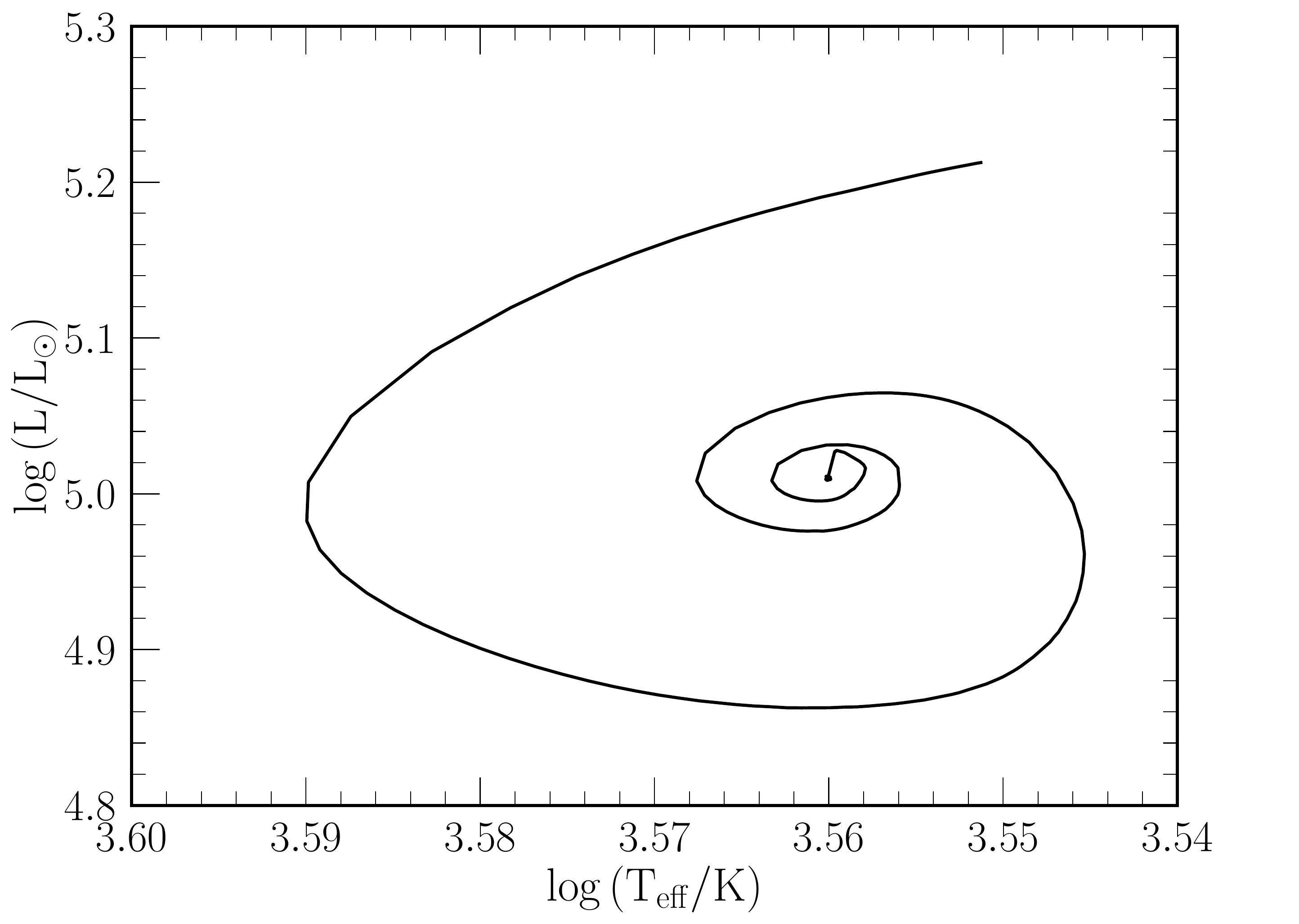}
    \caption{The surface temperature and luminosity for a our default 5\msun{} \tzo{} during the dynamical pulsations. This plot shows $\approx180$ years of evolution.}
    \label{fig:pulse_spiral}
\end{figure}

\begin{figure}
    \centering
    \includegraphics[width=\linewidth]{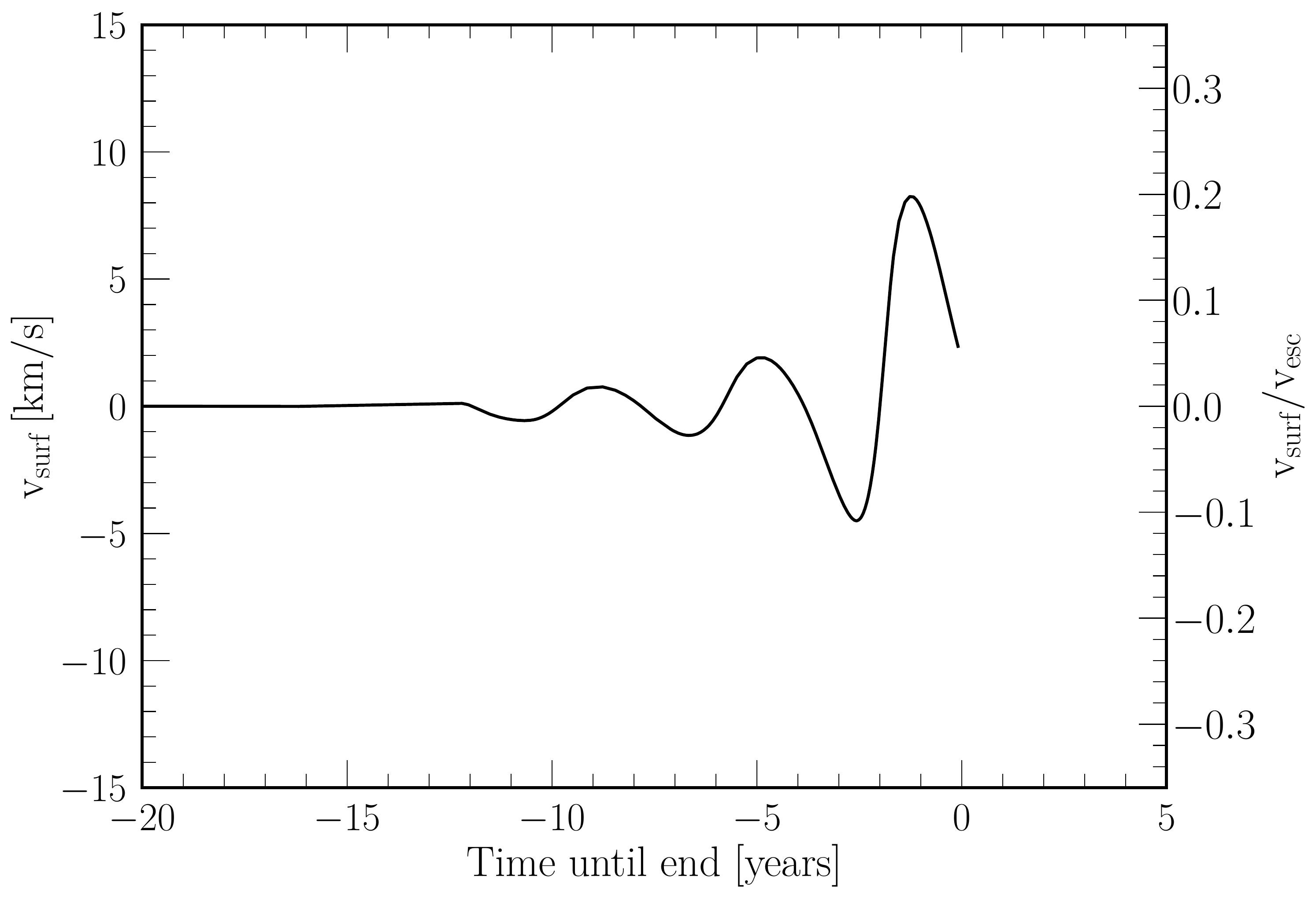}
    \caption{The radial velocity of the surface layers of our $5\msun{}$ default \tzo{} over time. The right-hand axis shows the approximate fraction of the escape velocity that the material reaches.
    The x-axis is the time until we can no longer follow the evolution with \MESA.}
    \label{fig:pulse_vsurf}
\end{figure}

Figure~\ref{fig:pulse_spiral} shows the HRD of our 5\msun{} default \tzo{} during this dynamical phase. Note this is only a representative plot, the exact shape of the spiral and number of cycles we can follow depends sensitively on the input physics and numerical resolution. We can see that as the star evolves the change in surface temperature and luminosity increases with each additional cycle.

Figure~\ref{fig:pulse_vsurf} shows the radial velocity of the surface layers of our default 5\msun{} \tzo{}. As the \tzo{} evolves the surface layers can reach velocities of $\approx\,10\,\kms$, and during the final contraction phase becomes supersonic. These velocities are of a similar magnitude to those found in \citet{yoon10} for the evolution of massive stars with pulsation-driven superwinds. This suggests that we should expect pulsation-driven mass loss in a \tzo. \citet{yoon10} finds the mass loss rate may increase up to $10^{-2}\,\msunyr$. These RSG pulsations are resolved when the timestep of our models drops significantly below the pulsation period. This occurs in our \tzos{} when the nuclear burning rate spikes, as seen in Figure \ref{fig:pulse_rates}, but can occur at earlier times if we artificially enforce a short timestep. If the mass loss rates are at the upper end of those predicted by \citet{yoon10} then this implies a lifetime of $\approx$ 100--1000 years. However, our wind mass loss is based on that of \citet{vanloon05} which is based on the observations of RSG's and thus has this mass-loss built into the time averaged mass-loss rates. More work is needed to understand the wind/pulsational mass loss rates in RSG's, as this will set the lifetime of a \tzo.

\section{Nucleosynthesis}\label{sec:nuc}

\begin{figure*}
    \centering
    \subfigure[Variations in \yinit]{\label{fig:big_net_y}\includegraphics[width=0.475\linewidth]{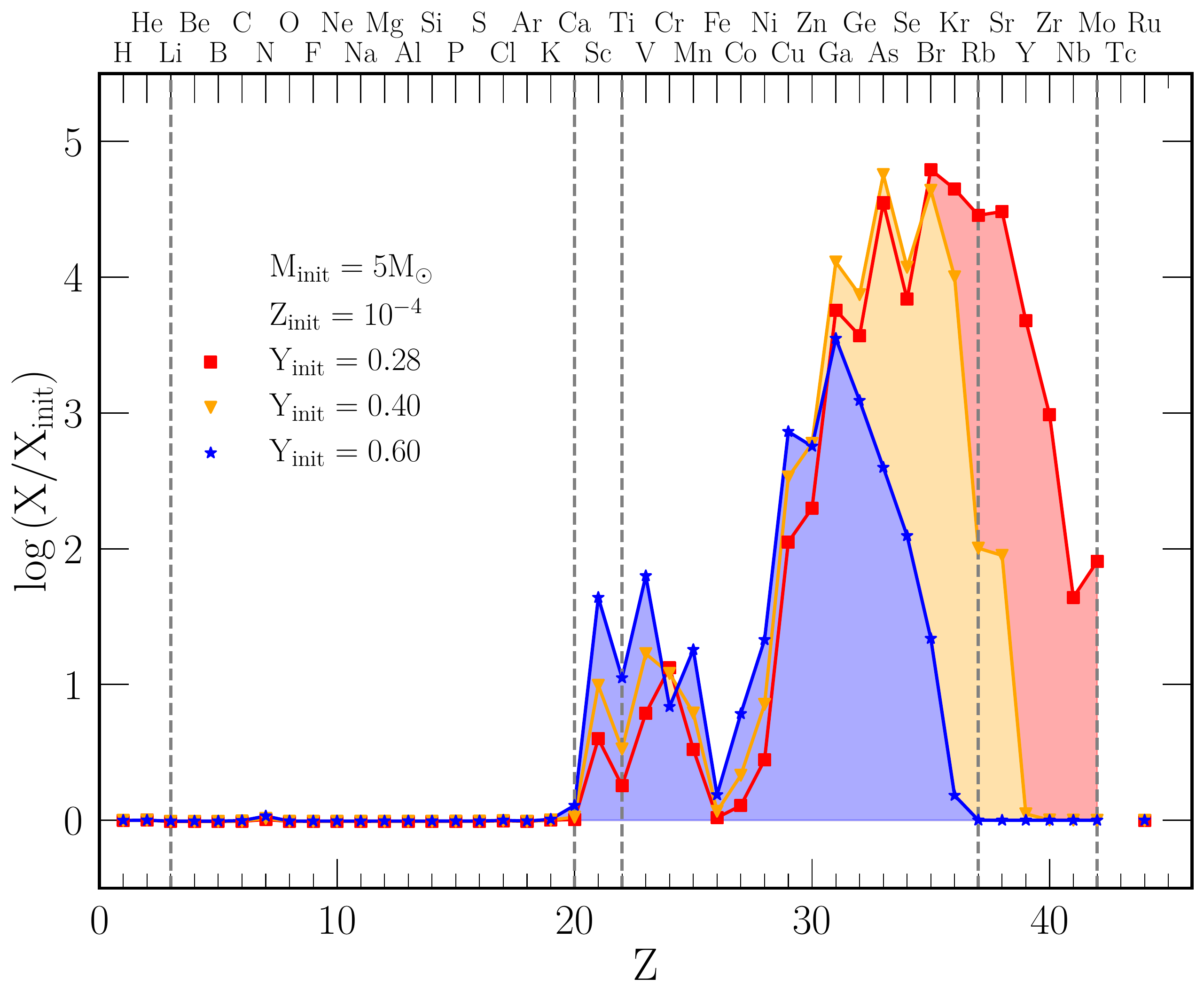}}
    \subfigure[Variations in \minit]{\label{fig:big_net_m}\includegraphics[width=0.475\linewidth]{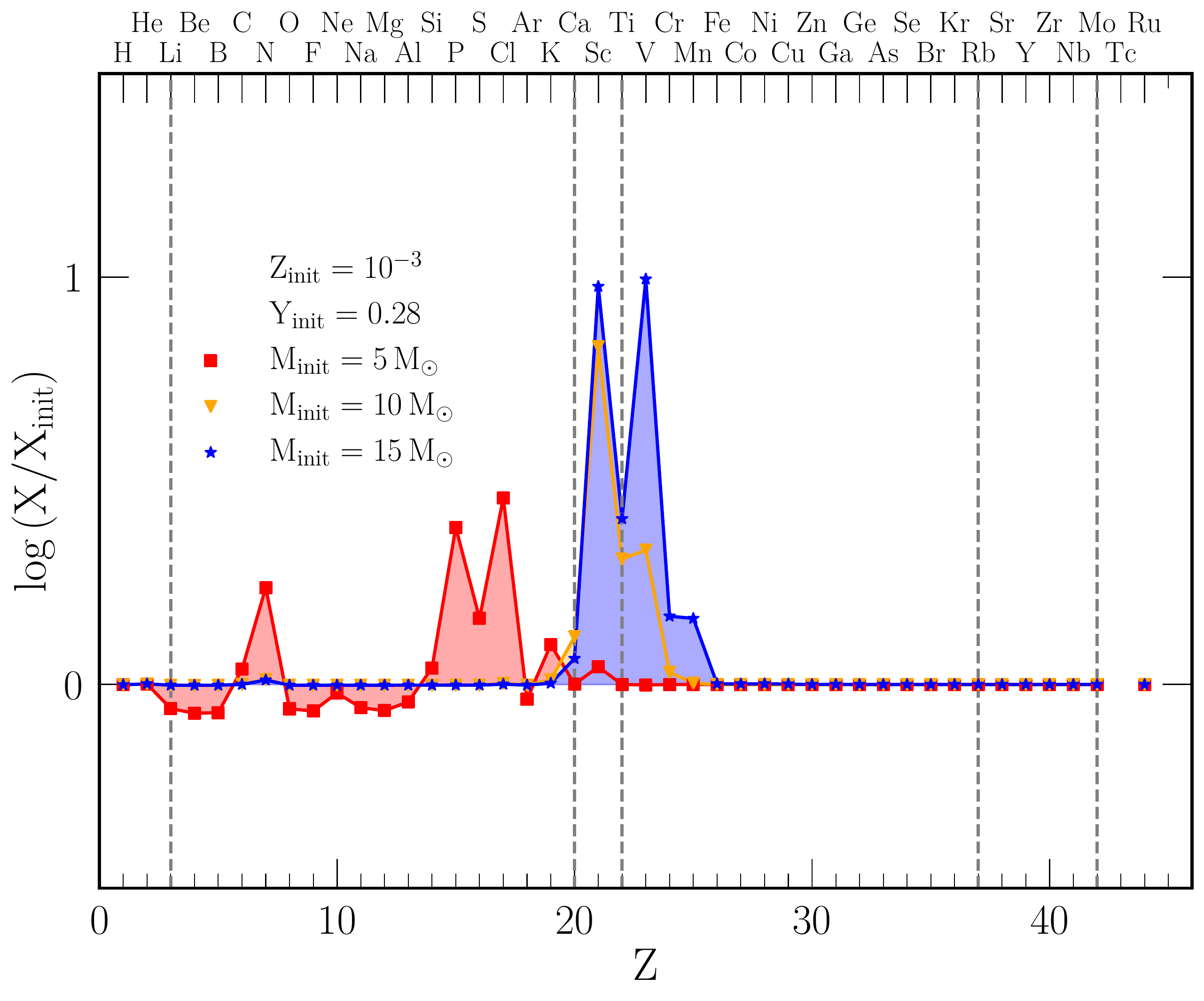}}

    \caption{The surface composition relative to the initial composition at 10,000 years post-\tzo{} formation. Left panel: The composition as a function of of the proton number Z for a 5\msun{} \tzo{} and $\zinit=10^{-4}$ with $\yinit=0.28$ (red square), $\yinit=0.4$ (orange triangle), $\yinit=0.6$ (blue star). Right panel: The composition as a function of of the proton number Z for $\yinit=0.28$ and $\zinit=10^{-3}$, with $\minit=5\msun{}$ (red square), $\minit=10\msun{}$ (orange triangle), and $\minit=15\msun{}$ (blue star).
     All models were evolved with a fully coupled \nisos{} isotope nuclear network.
    Vertical lines mark elements that may be useful for detecting \tzos. Note the change in the y-scale between panels.
    Data tables are available in the online Zenodo material with the time evolution of the composition.
    }
    \label{fig:big_net}
\end{figure*}

Figure~\ref{fig:big_net} shows the surface composition relative to the initial composition for \tzos{} evolved with  a fully coupled \nisos{} isotope nuclear network at $\approx10,000$ years after formation. Figure~\ref{fig:big_net_y} shows the composition as \yinit{} is varied, for a fixed $\minit=5\msun$ and $\zinit=10^{-4}$. Figure~\ref{fig:big_net_m} shows the composition as \minit{} is varied for fixed $\yinit=0.28$ and $\zinit=10^{-3}$.
A HRD for the models in Figure~\ref{fig:big_net_y} can be found in Appendix~\ref{sec:others}. We include all isotopes produced in our models, without taking into account any radioactive decay. A comparison with the results of \citetalias{cannon93} can be found in Appendix \ref{sec:other_nets}. We note, as a word of caution, that this abundance pattern is sensitive to the choice of the initial composition, as the initial metals act as seed nuclei for the \texttt{rp}-burning.

For the $\zinit=10^{-4}$ we find that for $Z<20$, there is no enhancement relative to the initial composition. For $21 \leq Z \leq 25 $ (Sc to Mn) there is an enhancement that increases as \yinit{} increases, while both Fe and Ca are not enhanced. For higher atomic numbers the enhancement relative to the starting composition becomes larger, up to $\approx10^5$ their starting values. As \yinit{} increases the most enhanced element decreases in atomic number, for $\yinit=0.28$ it is Br, $\yinit=0.4$ it is As, and $\yinit=0.6$ it is Ga. After this peak element, there is a rapid decline in the production of heavier elements.  Mo is only enhanced in our \yinit=0.28 model and is thus not a good element for determining a \tzo{} status.

These differences are due to the knee temperature decreasing as \yinit{} increases. This lowers the maximum mass of an element that can be produced in the \texttt{irp}-process. We reconfirm the previous findings of \citet{biehle91} and show that we can have significant enhancement of the elements Rb, Sr, Y, Zr and Mo but this depends sensitivity on the initial He composition. We find that models with very high initial helium fractions can lack Rb. Thus there could be a population of helium-rich \tzos{} without Rb, possibly explaining the difficulty in confirming the detection of a \tzo{} with the Rb mass fractions alone.

For the $\zinit=10^{-3}$ models we see almost no enhancement of metals at all. There is a slight enhancement of N due to CNO burning as well as an enhancement of S--Ar for the $\minit=5\msun$ model, and enhancements in Ca--Mn for the $\minit=10\msun$ and $15\msun$ models.
However there is no production of elements with $Z>25$. This is due to the higher opacity at the knee, lowering the knee luminosity, which lowers the knee temperature.  Thus, at metallicities comparable to the SMC, LMC, or MW, \tzos{} are unlikely to be distinguishable from non-\tzo{} stars based on their surface composition alone of elements such as Rb or Mo.

Table \ref{table:comp_ratio} shows the surface mass-fraction ratios for a selected set of elements for the $\zinit=10^{-4}$ models. Also shown is a comparison with Model A of \citetalias{cannon93}. We can see that the estimates for Ni/Fe are similar for all initial helium fractions. Estimates for Rb/Ni and Rb/Fe depend on the initial helium fraction but can be consistent with \citetalias{cannon93} for $0.28<\yinit<0.4$ (\citetalias{cannon93} assumes \yinit=0.32). Finally, Mo/Fe is extremely different, with our models predicting very little absolute Mo, with the valued decreasing as \yinit{} increases. Other ratios including K/Ca and Ca/Fe are a factor 10 smaller than \citetalias{cannon93}.

\begin{figure}
    \centering
    \includegraphics[width=\linewidth]{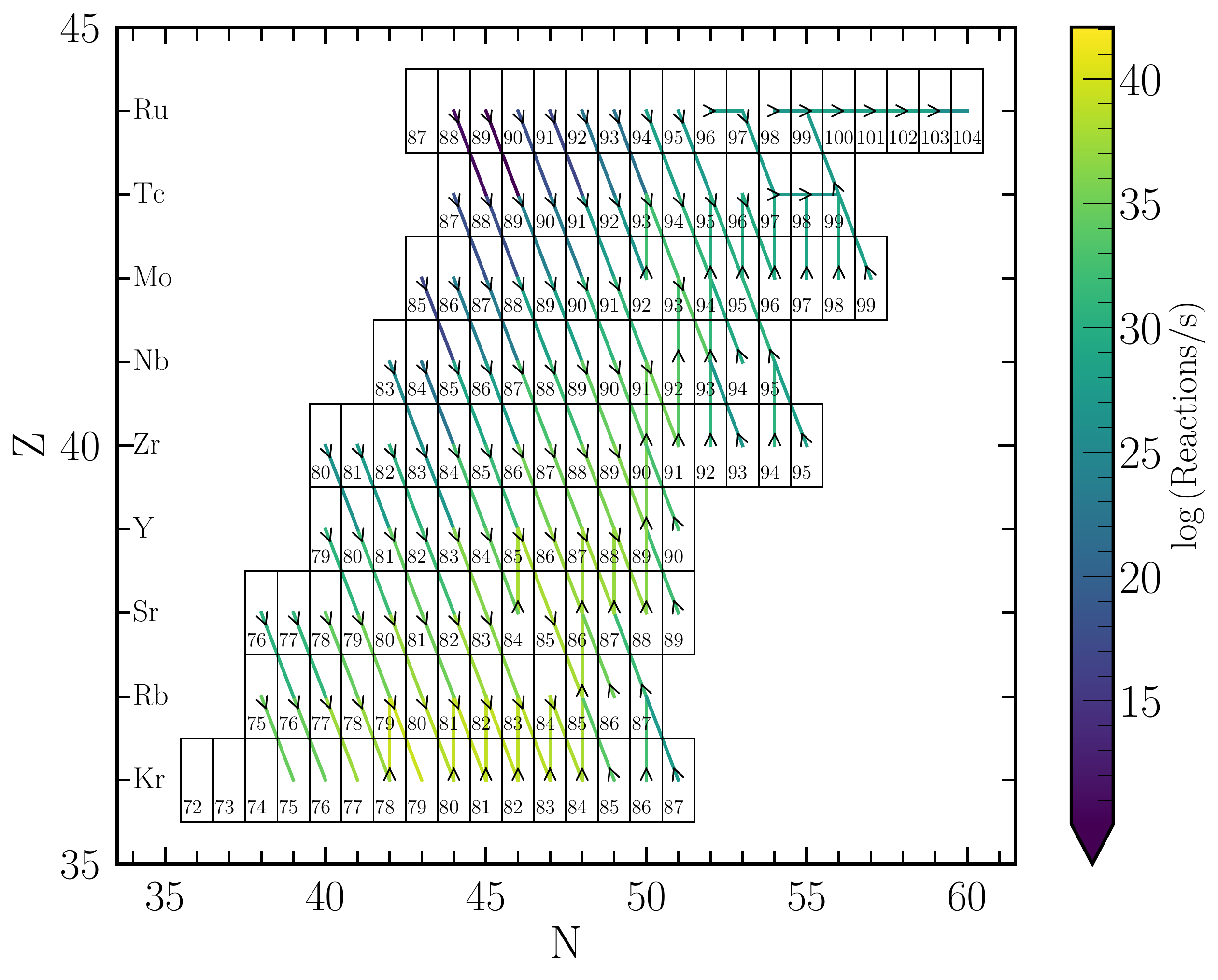}
    \caption{The total isotopic flow rate due to nuclear reactions for a selected set of isotopes in our default 5\msun{} \tzo{} evolved with an \nisos{} isotope nuclear network.
    The coloured lines show the logarithm of the net number of reactions per second for each isotope, showing only the most significant reaction, taking into account both forward and reverse reactions and summed over the entire \tzo{}.
    Arrows show the direction of flow. The model shown is $\approx 10,000$ years post-\tzo{} formation. Isotopes with no arrows have total reaction rates less than the lower limit of $10^{10}$ reactions/s. The atomic mass is quoted for each isotope in each box.}
    \label{fig:rate_flow}
\end{figure}

Figure~\ref{fig:rate_flow} shows how the isotopic composition will change over time, by showing the reaction flow for the dominant reaction for each isotope. Diagonal lines to the lower right are due to beta decays. There are insufficient free neutrons for the $(n,p)$ reactions to dominate as both \carbon[13] and \neon[22] have mass fractions $X\lesssim 10^{-17}$ at the base of the envelope. Vertical lines indicate a proton capture, while horizontal lines denote neutron captures. $\alpha$-captures are included in the network but are rarely significant in this mass range. Isotopes with no lines are included in the network but have all their reaction rates $<10^{10}$ reactions/s.

Isotopes are predominately decaying via beta-decays faster than proton captures can increase the atomic number of the isotope. Thus the flow is towards more neutron-rich isotopes, rather than more proton-rich isotopes. This prevents significant \texttt{rp}-burning from occurring and causes the limited amount of Mo shown in Figure~\ref{fig:big_net_y}. The maximum atomic number that is reached, before beta-decays outpace the proton captures will depend on the peak temperature reached by the \tzo{} at the base of the convection zone \citep{fisker08}. Thus, given Figure~\ref{fig:trho_int_mass}, the peak atomic number reached depends on the initial mass as well.

We have tested our choice of nuclear network in Appendix~\ref{sec:other_nets}.  Changing our nuclear network to one that closely matches that of \citetalias{cannon93}, has little effect on the results (see Appendix~\ref{sec:other_nets}). Though by extending to higher atomic numbers we can see a similar turn over in the mass fraction pattern in \citetalias{cannon93} as we see, except for \citetalias{cannon93} this occurs at higher atomic numbers.

\begin{table}
\centering
\begin{tabular}{|c|ccc|c|}
 \hline
 Element & $\yinit=0.28$ & $\yinit=0.40$ & $\yinit=0.60$ & \citetalias{cannon93} \\
 ratio &  & & & Model A \\
\hline
Rb/Ni & 2.18E+00 & 3.00E-03 & 9.94E-06 &  1.82E-01 \\
Rb/Fe & 3.38E-01 & 1.07E-03 & 8.04E-06 &  1.48E-01 \\
Li/Ca & 1.55E-04 & 1.49E-04 & 1.23E-04 &  \\
Li/K & 2.67E-03 & 2.67E-03 & 2.64E-03 &  \\
Mo/Fe & 3.99E-04 & 4.44E-06 & 3.36E-06 &  1.23E-01 \\
Ni/Fe & 1.55E-01 & 3.57E-01 & 8.08E-01 &  8.13E-01 \\
K/Ca & 5.80E-02 & 5.59E-02 & 4.64E-02 &  1.58E-01 \\
Ca/Fe & 4.91E-02 & 4.58E-02 & 4.23E-02 &  8.91E-01 \\
Rb/Zr & 1.76E+01 & 5.99E+01 & 5.97E-01 &  4.17E+00 \\
 \hline
\end{tabular}
\caption{Ratio of surface mass fractions for selected elements for our 5\msun{} \tzos{}, evolved with the large \nisos{} isotope nuclear network, at 10,000 years after \tzo{} formation at $\zinit=10^{-4}$. The final column contains the Model A data from \citetalias{cannon93} which only provides mass fractions for Carbon and heavier elements and thus lacks Lithium for comparison with.}
\label{table:comp_ratio}
\end{table}

Neutrino losses from our \tzos{} are $\approx 10^{37} \ergs$. This is dominated by the losses due to beta-decays, while the thermal neutrino losses are negligible. This is only a lower limit on the neutrino flux, as there may be additional neutrino emissions from the material below the knee which we do not model.
This neutrino flux is comparable to a solar-mass star at the tip of the red giant branch \citep{farag20}, and is unlikely to be detectable with current detectors \citep{patton17a,patton17b}.

\subsection{$^{44}\rm{Ti} \rm{O}_2$ and $^{44}\rm{Ti} \rm{O}$}

While the predicted unique nucleosynthetic signal of a \tzo{} has been used previously to make claims for the detection (or not) of a \tzo, it is not without controversy \citep{tout14}. Thus we propose a new signal which  provides a more constraining nucleosynthetic signal, namely the detection of molecules of TiO$_2$ and TiO containing the radioactive isotope \titanium[44].

\titanium[44] has a half life of $\approx60$ years \citep{audi03,ahmad06} and is usually found in the ejecta of core-collapse supernovae \citep{iyudin94}. Typical core-collapse supernovae have ejecta of $10^{-5}$--$10^{-4}\,\msun$ of \titanium[44] rich material \citep{magkotsios10}. This suggests that the detection of \titanium[44] in a \tzo{} could be the result of contamination from the supernovae that formed the NS initially. However, given its short half-life, unless we detect a \tzo{} shortly after the birth of the NS (when the SN remnant should still be visible) then the detection of \titanium[44] requires a continuous production site.

In non-\tzo{} stars, \titanium[44] is normally produced deep in the stellar core, via $\calcium[40](\alpha,\gamma)\titanium[44]$ during the late stages of stellar evolution and during explosive burning episodes \citep{timmes96}. Coupled with this, is that we need to be able to efficiently mix the \titanium[44]{} to the surface of the star, which is difficult to achieve unless the convective envelope penetrates deep into the core which is unexpected when $\calcium[40](\alpha,\gamma)\titanium[44]$ is active.

For reference we computed a $\mint=5\msun$ AGB star, $\mint=8\msun$ SAGB star and $\mint=10\msun$ massive star with $\zinit=0.00142$ ($\sim Z_{\odot}/10$).  For computational reasons we used a truncated version of our \nisos{} isotope nuclear network. We took our \nisos{} network and removed all elements heavier than Fe, as we are only interested in the Ti isotopes, which brings the total isotope count down to 172.

The 5\msun{} model is evolved until $\approx 40$ thermal pulses while the 8 and $10\msun$ models are evolved up to carbon ignition. We find the maximum surface mass fraction of \titanium[44]/\titanium[48] remains effectively zero (and always lower than the numerical tolerance imposed on the  nuclear network solver) throughout the evolution.

However, distinguishing different isotopes directly in a spectrum can be challenging. Thus we propose looking for isotopologues of $\rm{Ti}\rm{O}_2$ and $\rm{Ti}\rm{O}$. Our \tzos{} almost always have surface temperatures $\log\ (T_{\rm eff}/K)<3.6$, where $\rm{Ti}\rm{O}_2$ and $\rm{Ti}\rm{O}$ molecules are expected to form \citep{pavlenko20}, thus there should be a significant amount of $\rm{Ti}\rm{O}_2$ and $\rm{Ti}\rm{O}$ in the atmospheres of the \tzos.

The most common Ti isotope is \titanium[48], with contributions from \titanium[46-50] \citep{asplund09}. Detecting the different isotopologues has been achieved for these isotopes, due to a shift in the molecular lines as the mass of the molecules $\rm{Ti}\rm{O}_2$ and $\rm{Ti}\rm{O}$ change with the different isotopic compositions \citep{breier19}. The size of this shift is relative to the change in the molecular mass between the isotopologues \citep{herzberg50}. Thus a molecule containing \titanium[44] will have a larger shift in its molecular lines than the already detectable isotopologues containing \titanium[46-50] \citep{pavlenko20,serindag21}.  It may also be possible to use millimetre/submillimetre observations to detect $\titanium[44] \rm{O}$ given the detection of other TiO isotopologues \citep{kaminski13,lincowski16}.

\begin{figure}
    \centering
    \includegraphics[width=\linewidth]{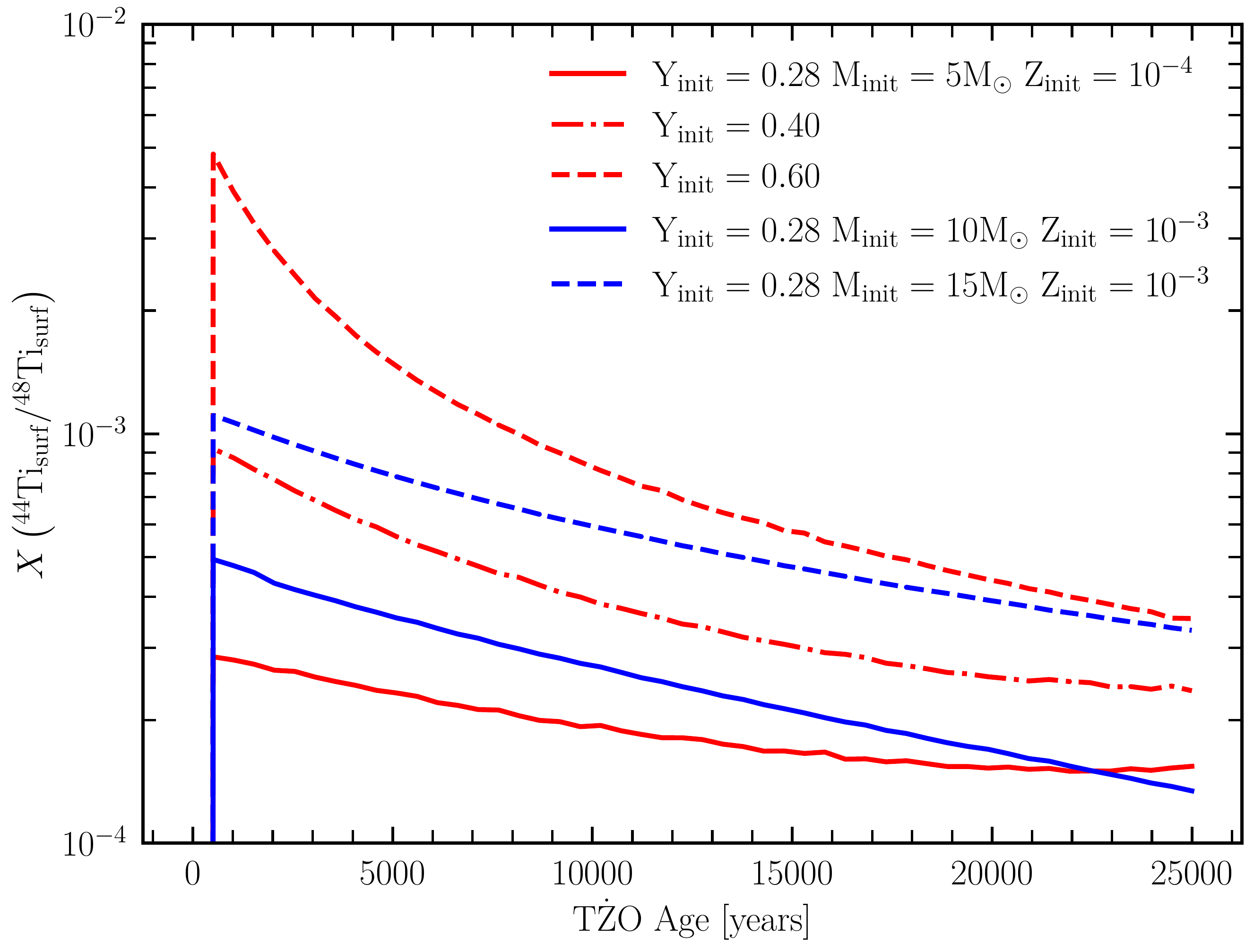}
    \caption{The ratio of the surface mass fraction of \titanium[44] to surface \titanium[48] over the \tzos{} lifetime.  In blue models with $\minit=5\msun$, and $\zinit=10^{-4}$ where solid line has $\yinit=0.28$, dot-dashed $\yinit=0.4$, and dashed $\yinit=0.6$. In green models with $\zinit=10^{-3}$ and $\yinit=0.28$, where the sold line has $\minit=10\msun$, and dashed has $\minit=15\msun$.
    Models were stopped at $\approx25,000$ years.}
    \label{fig:ti44_ratio}
\end{figure}

In our \tzo{} models \titanium[44] is produced in the CaScTi\footnote{While \MESA's \texttt{approx21} net does contain \titanium[44]{}, it is only produced via $\calcium[40](\alpha,\gamma)\titanium[44]$. Thus the \texttt{approx21} show significantly smaller amounts of \titanium[44]{} than when using our large nuclear network. } cycle via $\scandium[43]\left(p,\gamma\right)\titanium[44]$ \citep{fisker08}. The \titanium[44] is then mixed outwards before it can capture another proton. Figure~\ref{fig:ti44_ratio} shows the surface mass fraction of \titanium[44]/\titanium[48]. Depending on \yinit{}, \zinit{}, and \minit{} the ratios starts between $10^{-4}$--$10^{-2}$ and decreases with time. On long timescales, we can see that the ratio tends towards $X(\titanium[44]/\titanium[48])\approx10^{-4}$. This is due to increasing amounts of \titanium[48] (other stable Ti isotopes are also increasing with time), while the absolute amount of \titanium[44] remains approximately constant.
For the \tzos{} with $\zinit=10^{-3}$ the ratio decreases with \minit{}. A 5\msun{} \tzo{} with $\zinit=10^{-3}$ did not show any enrichment of \titanium[44]. These ratio's are much greater than the values found for the 5, 8, and 10\msun{} non-\tzo{} stars.

\begin{table*}
\centering
\begin{tabular}{ccccccccc}
 \hline
 \minit & \yinit & \zinit & \titanium[44] & \titanium[46] & \titanium[47] & \titanium[48] & \titanium[49] & \titanium[50] \\
\hline
5 & 0.28 & $10^{-4}$ & 3.85E-12 & 2.10E-09 & 2.76E-09 & 2.02E-08 & 6.50E-09 & 9.71E-10 \\
5 & 0.40 & $10^{-4}$ & 8.39E-11 & 1.03E-08 & 2.12E-08 & 1.01E-07 & 6.88E-08 & 9.96E-10 \\
5 & 0.60 & $10^{-4}$ & 1.31E-11 & 3.38E-09 & 5.66E-09 & 3.35E-08 & 1.71E-08 & 9.77E-10 \\
5 & 0.28 & $10^{-3}$ & 5.56E-16 & 1.43E-08 & 1.32E-08 & 1.34E-07 & 1.00E-08 & 9.86E-09 \\
10 & 0.28 & $10^{-3}$ & 5.46E-11 & 4.45E-08 & 4.57E-08 & 2.02E-07 & 6.47E-08 & 1.07E-08 \\
15 & 0.28 & $10^{-3}$ & 1.51E-10 & 3.96E-08 & 4.95E-08 & 2.55E-07 & 1.08E-07 & 1.02E-08 \\
\multicolumn{3}{c}{Sun} &  &2.23E-07 & 2.08E-07 & 2.15E-06 & 1.64E-07 & 1.64E-07 \\
5 &  & $Z_{\odot}/10$ & 3.05E-26 & 2.02E-08 & 1.85E-08 & 1.88E-07 & 1.44E-08 & 1.39E-08 \\
8 &  & $Z_{\odot}/10$ & 2.16E-98 & 2.05E-08 & 1.89E-08 & 1.91E-07 & 1.43E-08 & 1.41E-08 \\
10 &  & $Z_{\odot}/10$ & 2.57E-98 & 2.05E-08 & 1.89E-08 & 1.91E-07 & 1.43E-08 & 1.41E-08 \\
 \hline
\end{tabular}
\caption{Surface mass fractions for different Ti isotopes for our large nuclear network models. The composition is taken at $\approx 10,000$ years post \tzo{} formation. The 5, 8, and 10\msun{} non-\tzo{} stars are measured at the time of maximum surface \titanium[44]. Solar values from \citet{grevesse:98}. }
\label{table:ti_abun}
\end{table*}

Table \ref{table:ti_abun} shows the absolute mass fraction fractions of the Ti isotopes in our large nuclear network models. We can see that the amount of all Ti isotopes slightly increases with increasing initial helium fraction. There is however a factor $\sim10$ increase in \titanium[44] as \yinit{} increases.  Given the small absolute mass of \titanium[44], which is much less than that typically seen in supernovae ejecta \citep{magkotsios10}, it may be difficult to directly detect the gamma-ray from the decay of \titanium[44].

We note though that before this becomes a viable method it is likely we will need improved theoretical models of the molecular lines of $\titanium[44] \rm{O}_2$ and $\titanium[44] \rm{O}$, as to date experiments have concentrated on the more common and stable isotopes of titanium, namely 46--50 \citep{brunken08,breier19,mckemmish19,witsch21}.

\section{Suitability of model assumptions}\label{sec:assume}

\begin{figure*}
     \centering
     \subfigure[Central density]{\label{fig:log_l_rho}\includegraphics[width=0.49\linewidth]{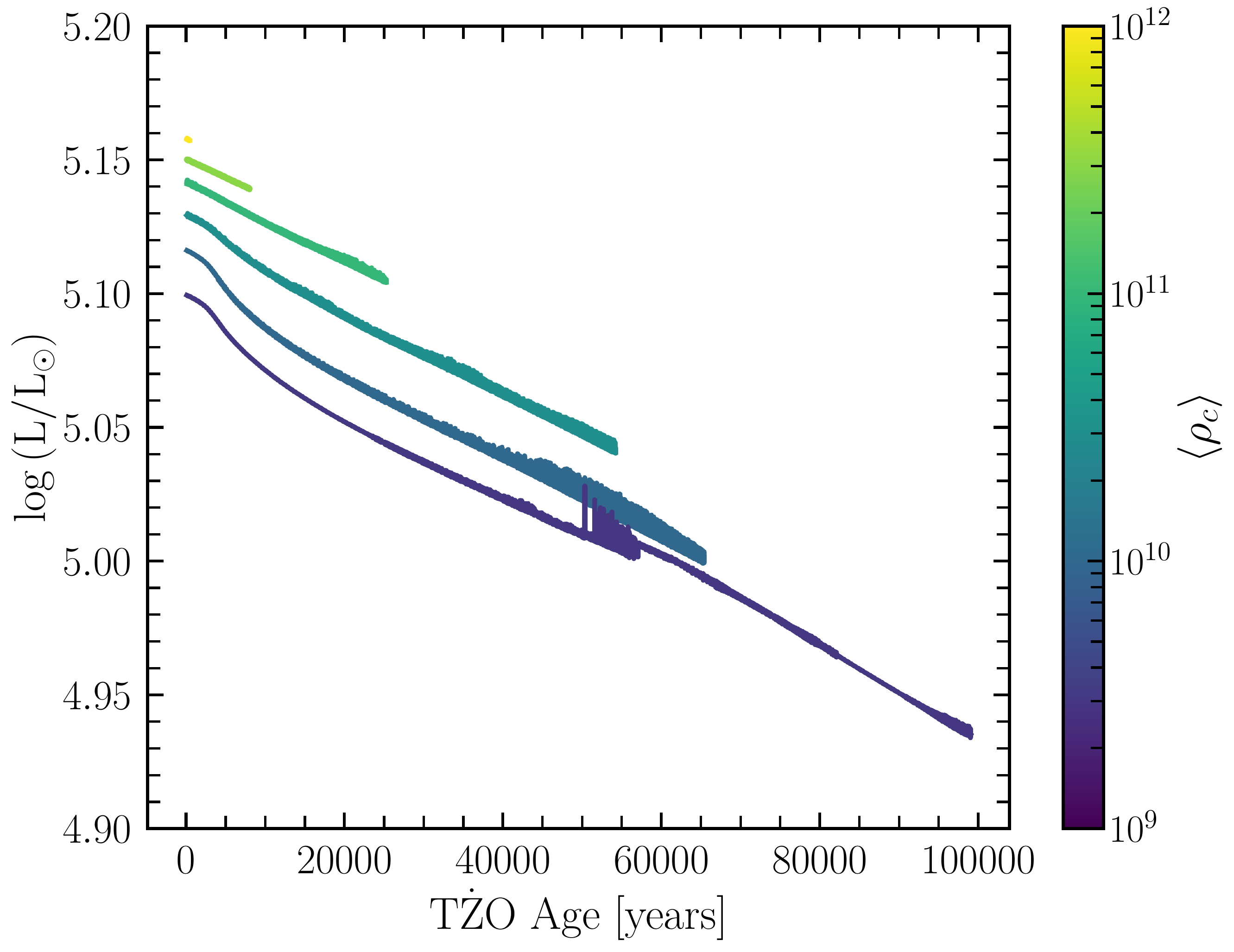}}
     \subfigure[Central luminosity]{\label{fig:logl_ledd}\includegraphics[width=0.49\linewidth]{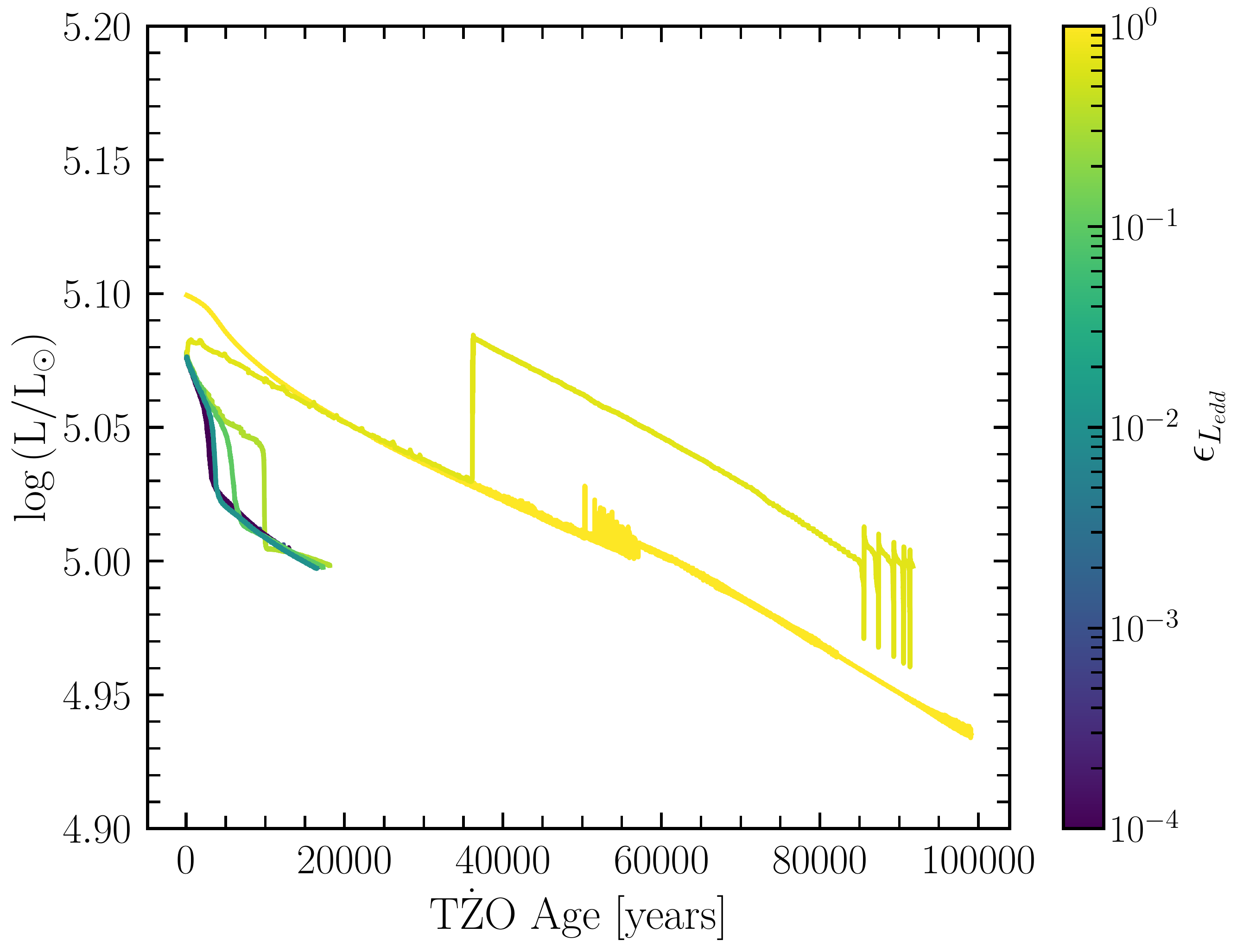}}

        \caption{ Panel a: The evolution of the surface luminosity as a function of the assumed average core density. Light colours denote higher core densities and smaller assumed NS radii. Panel b: The evolution of the surface luminosity as a function of the efficiency factor \eledd. Light colours denote higher efficiencies and thus the energy injected into the inner boundary is closer to \ledd.  Evolution was arbitrarily stopped when either the model reached 100,000 years, 100,000 timesteps, or when \MESA{} could no longer follow the evolution. }
        \label{fig:logl_params}
\end{figure*}

\begin{figure}
    \centering
    \includegraphics[width=\linewidth]{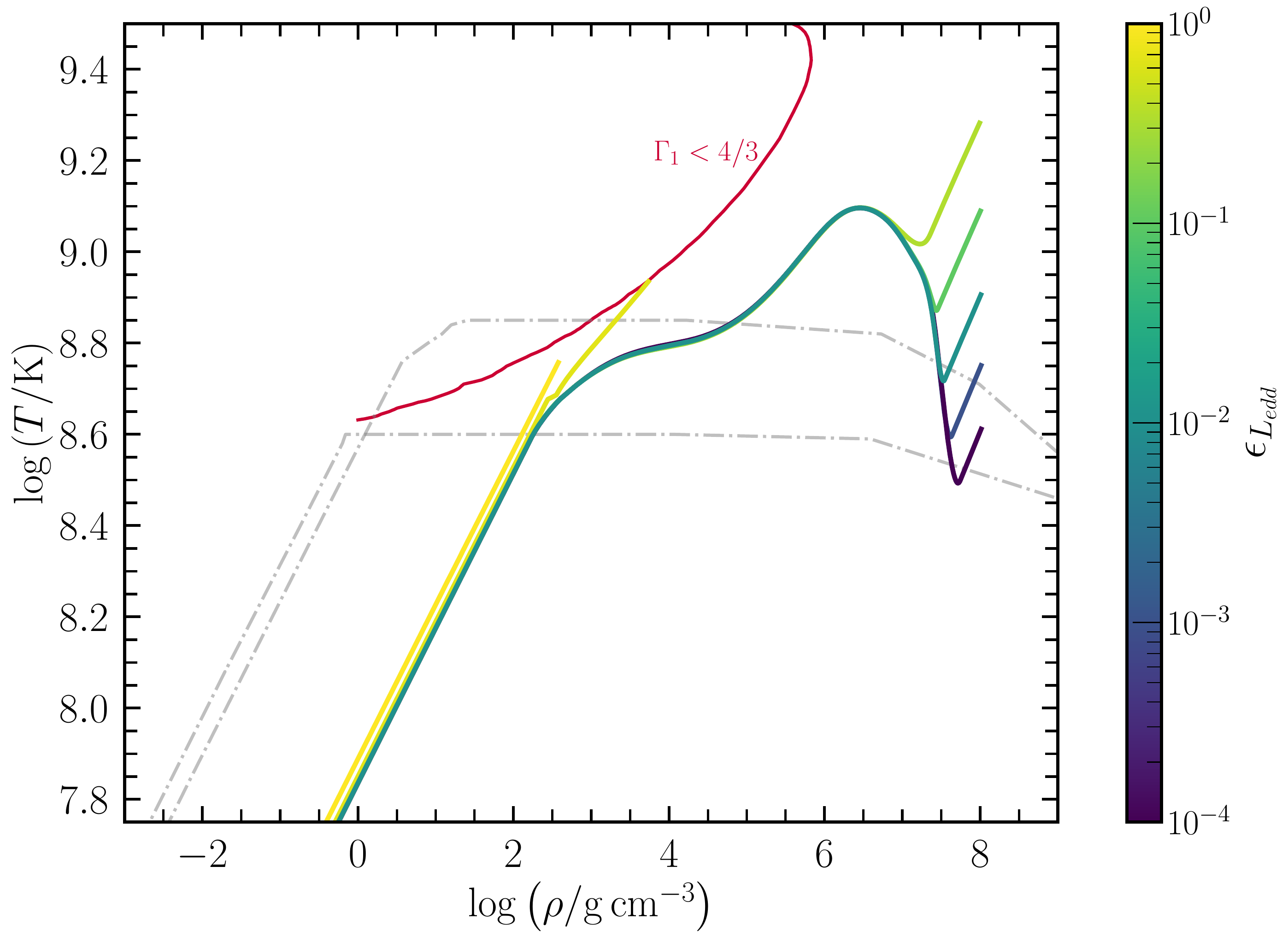}
    \caption{Top Panel: The temperature-density (T-$\rho$) profile for the models shown in Figure~\ref{fig:logl_ledd} at a time $\approx 10,000$ years post \tzo{} formation. The $\eledd = 0$ model did not reach 10,000 years and is thus not shown. Dash-dotted lines are the models of \citetalias{cannon92} while the red region is the pair-instability region.}
    \label{fig:logl_trho}
\end{figure}

The models presented in this work depend on two key approximations, that make the \tzo{} models computationally feasible. Without these approximations, these models would not be computationally viable. Both approximations deal with the fact that we can not model down to the surface of the NS while simultaneously including the RSG envelope. First, we assume a $\log\ \avgrhocgcm= 9.3$, which implies an effective $\rm{R_{NS}}\approx600\,\km$ and secondly we inject additional energy at the base of our models. We will now test those assumptions, as far as possible, to determine their effect on our results. For this test, we turn off the hydrodynamics, as it can cause numerical convergence issues.

Figure~\ref{fig:logl_params} shows the total luminosity of the \tzo{} as a function of the age since formation, for variations in the core density (\ref{fig:log_l_rho}), and the injected energy (\ref{fig:logl_ledd}). Firstly we can see that both sets of models evolve to lower luminosities over time. Figure~\ref{fig:log_l_rho} compares models with $\log \avgrhocgcm$ between $10^9$--$10^{12}\,\grampercc$ (this equates to effective NS radii between $\approx 90$--$900\,\km$). It shows that as the core density increases the total luminosity increases. This is due to a lower opacity at the knee, causing \lknee{} to increase. Figure~\ref{fig:logl_ledd} also shows that as the injected energy decreases, the surface luminosity also decreases. Thus, surprisingly, over the range of parameters explored here, the two approximations act with different signs but similar magnitudes, in terms of the total luminosity. Therefore they approximately cancel out. Comparing the surface metal mass fraction, the low \eledd{} models converge to values about a factor 2 greater than the $\eledd=1$, while the highest core density models have a factor 2 lower surface metal mass fractions. Again, our two assumptions are approximately cancelling each other out.

The large luminosity jumps leading to the models then evolving on a different luminosity track are due to the lack of hydrodynamics. In models with hydrodynamics included, those jumps occur at the time when the energy generated by nuclear burning spikes and the models would normally have been stopped due to large amplitude surface pulsations being resolved in the models.

In Figure~\ref{fig:logl_trho} we show the temperature-density profiles inside the \tzos{} as a function of \eledd{}. As \eledd{} decreases, less energy is injected into the model, therefore the \tzo{} must provide more of the energy itself. As the mass of the NS varies little in our models this is also equivalent to changing the opacity of the material at the base of the convection zone. The material below the knee then moves to higher temperatures and densities to provide the necessary energy via nuclear burning and mass accretion to support the star. The region below the knee does not evolve at a constant temperature, as previously found in \citet{thorne75,thorne77} and \citetalias{cannon92}. Instead as the temperature increases, the material avoids the pair-instability region by evolving around the instability region. The T--$\rho$ tracks converge for $\eledd \leq 0.1$, until $\log (\rho / \grampercc) \approx 6.0$.

At $\log (\rho / \grampercc) \approx 6.0$ the core begins evolving to cooler temperatures. At this temperature both hydrogen and helium are depleted in the material below the knee. With a very low mass fraction of carbon and oxygen, due to the low initial metallicity, and our use of the \texttt{approx21.net} nuclear network, the nuclear energy generation rate goes to 0 as there are no available nuclear reactions. The inner regions then begin to cool due to an increase in thermal neutrino losses as the density increases.

Finally, the increase in temperature at $\log (\rho / \grampercc) \approx 8.0$ is due to a new convection zone setting in near the inner boundary of the model. This is due to the (small) amount of energy that is still being injected into the model. Models with less injected energy cool further and reach higher densities before showing this uptick.

\section{Final fate}\label{sec:fate}

Once the large amplitude surface pulsations begin the computational timescale decreases and we begin evolving the models on the pulsation timescale. This leads to timesteps of order $10^2$--$10^3$ seconds, which is much smaller than the $10^8$ seconds we take during the normal phase of a \tzos{} evolution and thus is infeasible to evolve over longer time frames. Numerical convergence issues also occur as shocks form in the outer envelope, preventing the models from being evolved for more than a few tens of years at this point.

Thus we can only speculate what happens next. The \tzos{} are undergoing large pulsations, and in other $\epsilon$-mechanism pulsators this is expected to lead to pulsation-driven mass loss \citep{baraffe01,nakauchi20} or, as in normal RSGs, pulsations may lead to pulsational driven superwinds \citep{yoon10}. This could rapidly decrease the envelope mass over $\approx100$ years, leaving behind a bare NS. Or perhaps the \tzo{} will undergo a neutrino runaway when nuclear burning ceases and energy losses from neutrinos cause the \tzo{} to collapse \citep{Podsiadlowski95}, leaving a BH behind, though this seems less likely as our models avoid the pair-instability region.

If the envelope was entirely ejected from the NS, it seems unlikely that the NS mass will have increased significantly enough to be visible as a higher mass NS. The accretion rate onto the NS is $\approx 10^{-9}$--$10^{-8}\,\msunyr$. In our models, this leads to the NS gaining at most $\approx0.002\,\msun$ during its evolution. The NS may also be spun down by braking between the NS's magnetic field and the envelope \citep{liu05}. In this scenario the NS will become a slow spinning NS inside a slow moving CSM that appears like a supernovae remnant, that has been proposed for RCW 103 \citep{liu05}.

The other option is that the \tzo{} runs out of CNO material, which provides the bulk of the nuclear energy generated \citep{biehle94}. In this case, the envelope may collapse onto the NS forming a BH, possibly leading to a transient event \citep{moriya18,moriya21}. In a 5\msun{} \tzo{} model the total mass of CNO elements decreases at a rate of $\approx 10^{-10}\,\msunyr$. With an initial mass of $\approx10^{-5}\,\msun$ of CNO elements (for $\zinit=10^{-5}$), this gives an upper limit of $\approx10^6$ years, assuming the burn rate continues at the same rate. This is comparable to the lifetime given assuming a steady state wind mass loss.

If the NS inside the \tzo{} did collapse into a BH and form a transient event, this may look like a type IIn SN. The pre-SN pulsations likely removed 1--10\msun{} of material in the 100--1000 years before the final collapse. At collapse it is then likely that there is only a few solar masses, at most, of H-rich envelope left. Our default 5\msun{} \tzo{} has only a $\approx2.5\msun$ envelope left before the pulsations begin, while at 20\msun{} has $\approx5.5\msun$ envelope before the pulsations begin. We would not expect any significant \nickel[56]{} production in the event, and as the \tzo{} was fully mixed there would not be any observed change in the composition at late times. To power a SN transient event would require it either to be entirely powered by CSM interaction, energy released as the NS collapses into a BH, or if the BH generates a jet from additional mass accretion \citep{fryer96,qin98,fryer14}.

\section{Discussion}\label{sec:discuss}

In this work, we have been agnostic as to how a \tzo{} formed, whether it was a direct impact, common envelope merger, or dynamical merger. Each of the formation pathways may impact the resulting \tzo. Common-envelope mergers require that the companion star is evolved, with a helium-rich core and an envelope that is increasing in radius. While direct impacts and dynamical mergers are less sensitive to the companion star's evolutionary state, at least regarding when a merger can occur. Merging at different points in the star's pre-\tzo{} life will lead to a different initial metal fraction than our assumed solar-scaled metal distribution, and limits what values of \yinit{} are possible. There could be differences in whether or not a merger results in a \tzo{} or instead causes enhanced mass loss, leaving behind a tight binary with an NS, thus failed \tzo's are progenitors for double NS systems (DNS). Detections of \tzos{} could place constraints on the uncertain merger rate of DNS systems as currently being probed by gravitational wave detections.

We assumed that the starting point for a \tzo{} can be approximated as a normal star, which then becomes fully convective. However, it is possible that the companion was the accretor in a binary system. The material it gains will be enriched by the nuclear burning in the donor \citep{farmer21,farmer23a} and the internal structure of the accretor will be altered by the mass accretion \citep{renzo21,renzo22}. This may lead to possible different outcomes during the merger, which may include additional mass loss, changes in the probability of a successful \tzo{} formation, and to initially enrich the \tzo{} in additional metals.

It was argued in \citetalias{cannon93} that it was necessary to use a two-stream convection model to follow the nucleosynthesis. Here the composition is tracked for material that is convectivly moving towards the knee and away from the knee. As material moves towards the NS it undergoes proton captures, before mixing away from the knee, where it can then undergo beta decays, before being mixed back towards the knee. As the burning timescale is comparable to the mixing timescale at the base of the envelope the chemical composition of material flowing up/down may be significantly different. \MESA\ takes assumes a diffusion equation for chemical transport. The effect of having a diffusion model is that we likely over predict the production of heavy nuclei. In the \MESA{} models some heavy nuclei will be able stay near the base of the envelope, where the temperatures are highest, for much longer than they would if they where being mixed outwards on a timescale similar to their burning timescale. Thus they can undergo additional proton captures and produce even heavier nuclei. This may suggest we are even less likely to see a detectable nucleosynthetic signal in \tzos{} in the local Universe.

\section{Conclusions}\label{sec:conc}

In this work, we have computed the first set of \MESA{} \tzos{}. We did this by adjusting the inner boundary of the model to approximate the NS at the centre of each \tzo{}.  We have then followed the evolution of the \tzos{}. We have also explored in detail the pulsation periods with the \GYRE{} stellar oscillation instrument, and computed detailed nucleosynthetic signatures with a large fully-coupled \nisos{} isotope nuclear network. Our results can be summarised as follows:

\begin{itemize}
    \item We find that \tzos{} evolve to lower luminosities and lower temperatures during their lifetime. We have also expanded the range of possible locations for \tzos{} to be between $\logt \approx 3.47$--$3.6$ and $\logl \approx 5.0$--$5.5$.

    \item We do not find a gap in the parameter space where models can not exist. This is because our models are denser than previously predicted, which prevents models from evolving into the pair-instability region.

    \item We have computed the pulsation periods of our \tzos{} and find the longest periods to be $\approx 250, 500$ and 1000-2000 days.

    \item If HV 2112 is a \tzo{}, we predict there should be a currently undetected 1500--3000 day pulsation period. If detected this will also imply $\mlta\approx3$ in the envelopes of \tzos{}.

    \item If VX Sgr were a \tzo{} and its $\sim28,000$ day pulsation period were real, this would imply it is a very helium enriched \tzo{}. 

    \item Based on the measured pulsation periods of HV 11417 we would infer it to be a massive, but very young \tzo. However this is inconsistent with the measured luminosity of HV 11417. Thus we rule out HV 11417 as a \tzo{}.

    \item Our results and the predicted lifetimes depend strongly on the mass-loss rates due to RSG pulsations. The lifetime of the \tzo{} will depend on the total mass-loss rate experienced by the \tzo{}. If RSG pulsations remove significant amounts of material then the lifetime of a \tzo{} may only be 100--1000 years.

    \item Assuming the RSGs do not experience significantly higher mass loss than we assume, then we estimate a lifetime in the range of $10^4$--$10^5$ years. Contrary to non-\tzo{} stars, the higher the initial mass of the \tzo{} the longer it lives.

    \item We have computed several of our models with a large \nisos{} isotope fully-coupled nuclear network. We reconfirm the previous findings that Rb, Sr, Y, Zr can be enhanced due to the \texttt{irp}-burning. However, the level of enhancement is sensitive to the initial composition of the \tzos{}.

    \item At higher initial metallicities \tzos{} do not show any metal enrichment due to a lower knee temperature, caused by the increasing opacity at the knee as the metallicity increases. Thus in the local Universe \tzos{} may not be distinguishable from non-\tzos{} based on their nucleosynthesis alone.

    \item We propose a new observational signal, that of molecules containing \titanium[44]. Due to the high-temperature burning and fully convective envelope of the \tzo{}, \titanium[44] can be mixed to the surface before it decays. There in the cooler envelope, it can form \titanium[44]O$_{2}$ and \titanium[44]O, which could be detectable due to the shift in their molecular lines compared to stable Ti-containing molecules.

    \item \tzos{} represent a class of stars that are exceptional tests of the numerical capabilities of a stellar evolution code. This work has led to many improvements and code fixes in the \MESA{} stellar evolution code that have applications far outside that of \tzos{}.
\end{itemize}

\section*{Acknowledgements}

We acknowledge helpful discussions with T.~Maccarone, F.~Timmes, B.~Paxton, R.~Smolec, J.~Schwab, A.~Jermyn, R.~Townsend.
This work has been supported by the following grants at some point in time; NASA under TCAN grant NNX14AB53G (PI F.~Timmes), NSF under SI2 grant 1339600 (PI F.~Timmes), the Netherlands Organization for Scientiﬁc Research (NWO) through a top module 2 grant with project number 614.001.501 (PI S.E.~de Mink).
Support for this work was provided by NASA through the NASA Hubble Fellowship Program grant \#HST-HF2-51457.001-A awarded by the Space Telescope Science Institute, which is operated by the Association of Universities for Research in Astronomy, Inc., for NASA, under contract NAS5-26555.
This work was also supported by the Cost Action Program ChETEC CA16117.
This research was supported by the Munich Institute for Astro-, Particle and BioPhysics (MIAPbP) which is funded by the Deutsche Forschungsgemeinschaft (DFG, German Research Foundation) under Germany's Excellence Strategy – EXC-2094 – 390783311. This research was supported in part by the National Science Foundation under Grant No. NSF PHY-1748958.
This research has made use of NASA's Astrophysics Data System.

\section*{Data Availability}

All input files and all output data is made available at \url{https://doi.org/10.5281/zenodo.4534425}.



\bibliographystyle{mnras}
\bibliography{tzo} 



\appendix

\section{Sensitivity to other parameter choices}\label{sec:others}

\begin{figure*}
     \centering
     \subfigure[Wind efficiency]{\label{fig:hr_wind}\includegraphics[width=0.45\linewidth]{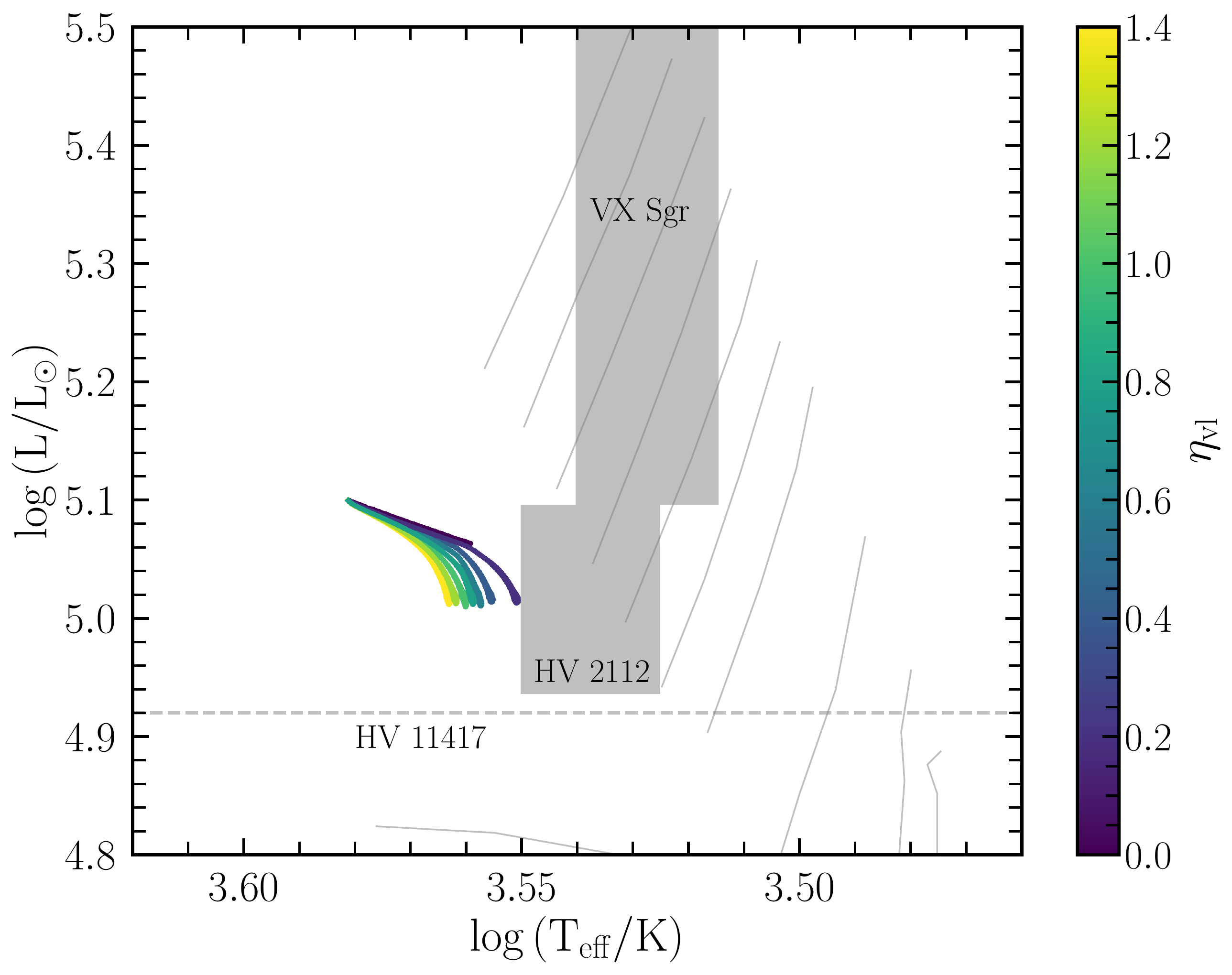}}
     \subfigure[Mixing Length alpha]{\label{fig:hr_mlt}\includegraphics[width=0.45\linewidth]{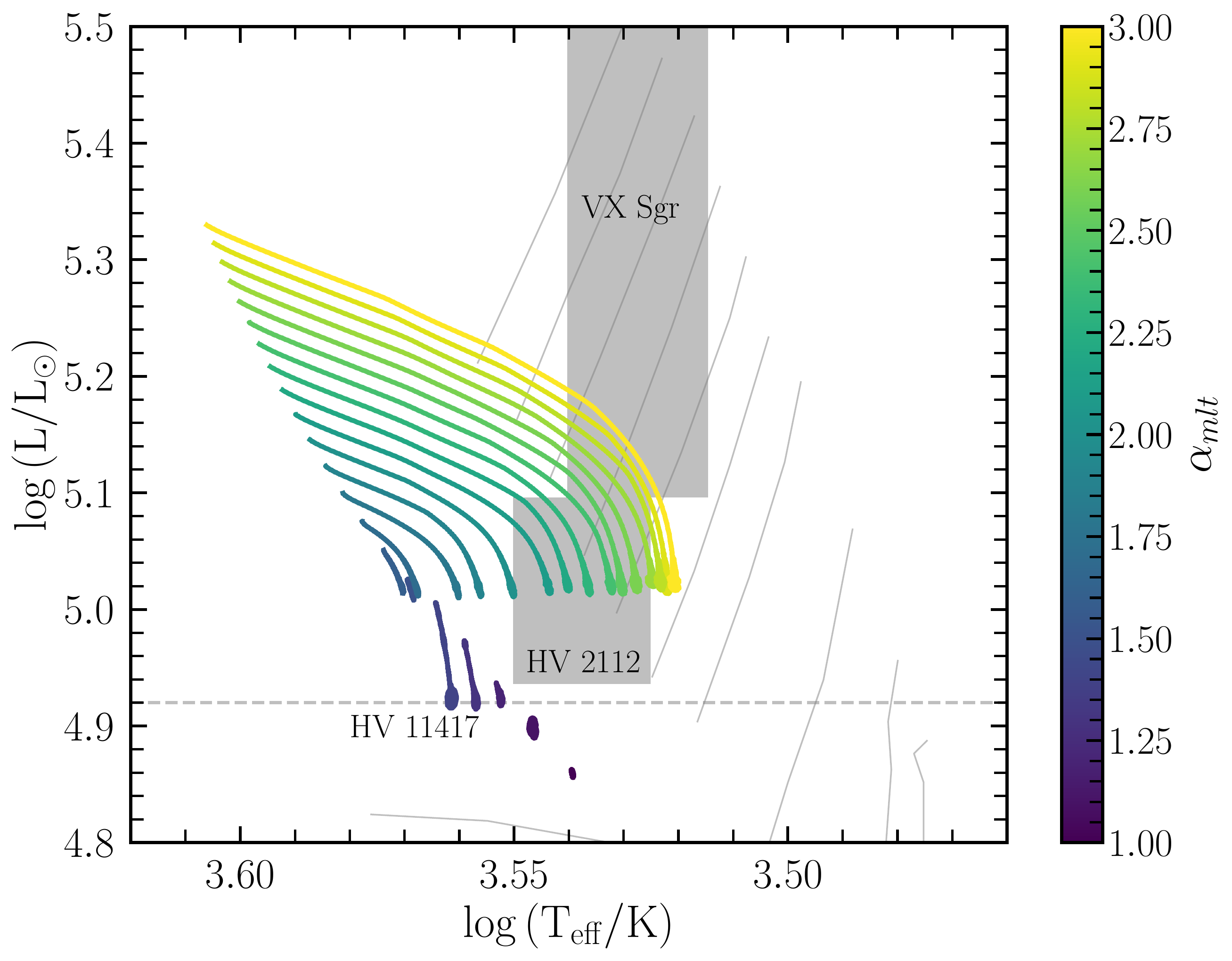}}

     \subfigure[Initial Mass NS]{\label{fig:hr_mns}\includegraphics[width=0.45\linewidth]{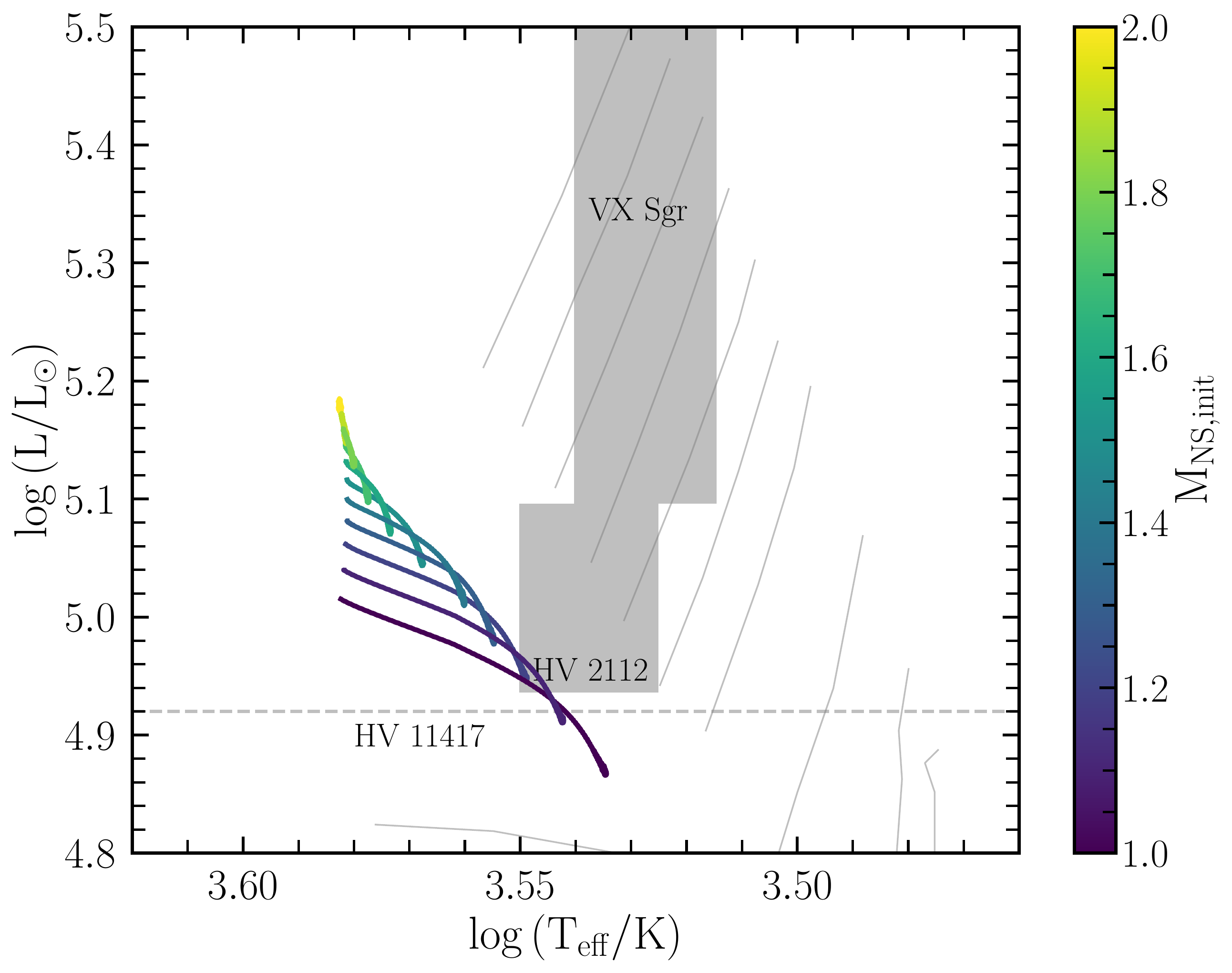}}
     \subfigure[NS accretion rate]{\label{fig:hr_ns_acc}\includegraphics[width=0.45\linewidth]{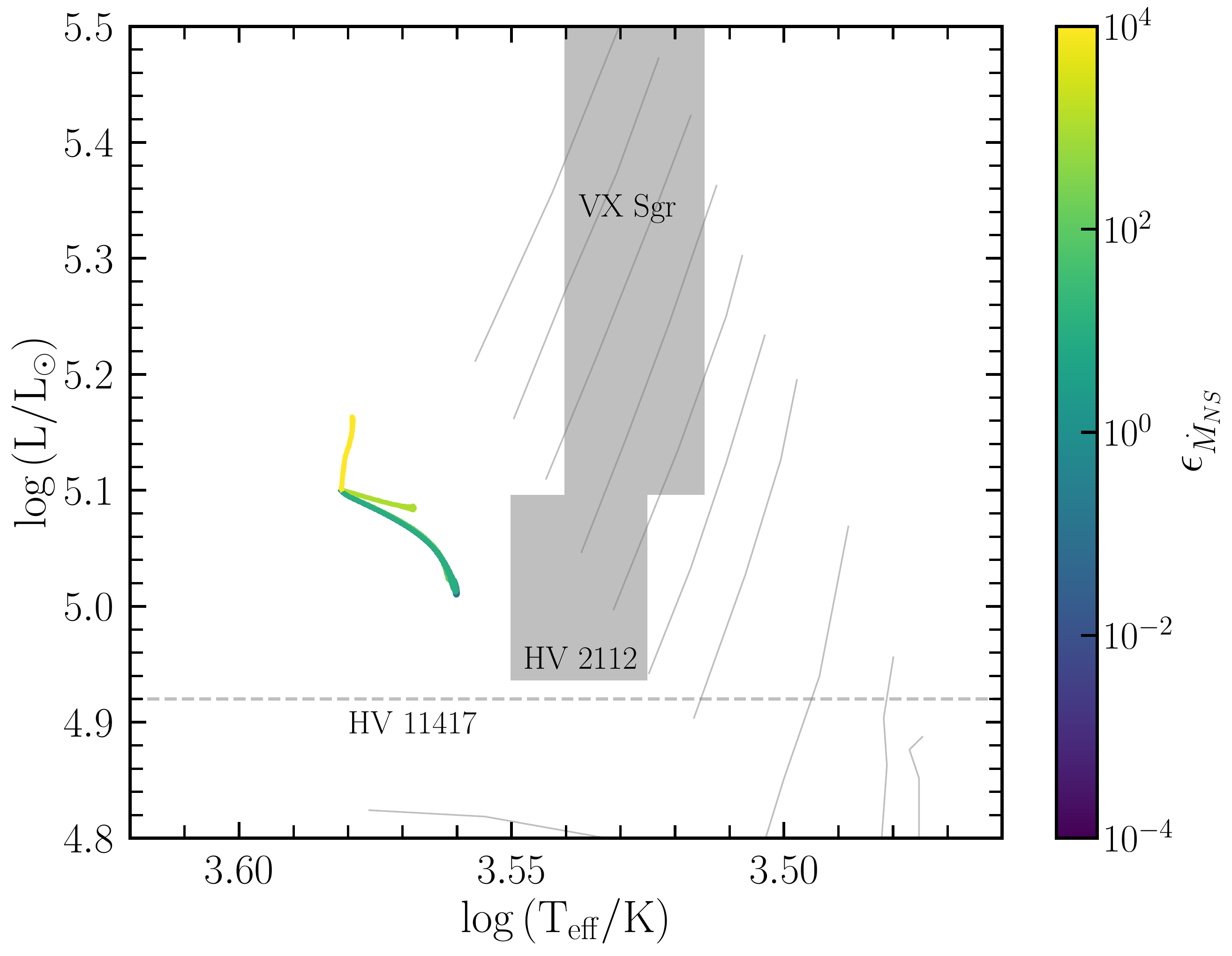}}

        \caption{HRDs for variations in our \tzo{} modelling assumptions. Grey lines and grey boxes have the same meaning as in Figure~\ref{fig:hrd_int_mass}. All models are taken as 5\msun{} with a $\mns=1.4\,\msun$, and use our default parameters except for those being varied in each panel. Panel a: variations in the wind mass loss efficiency, panel b: variations in the mixing length alpha, panel c: variations in the mass of the NS, and panel d: variations in the accretion rate onto the NS. }
        \label{fig:hr_params}
\end{figure*}

\begin{figure}
     \centering
     \includegraphics[width=\linewidth]{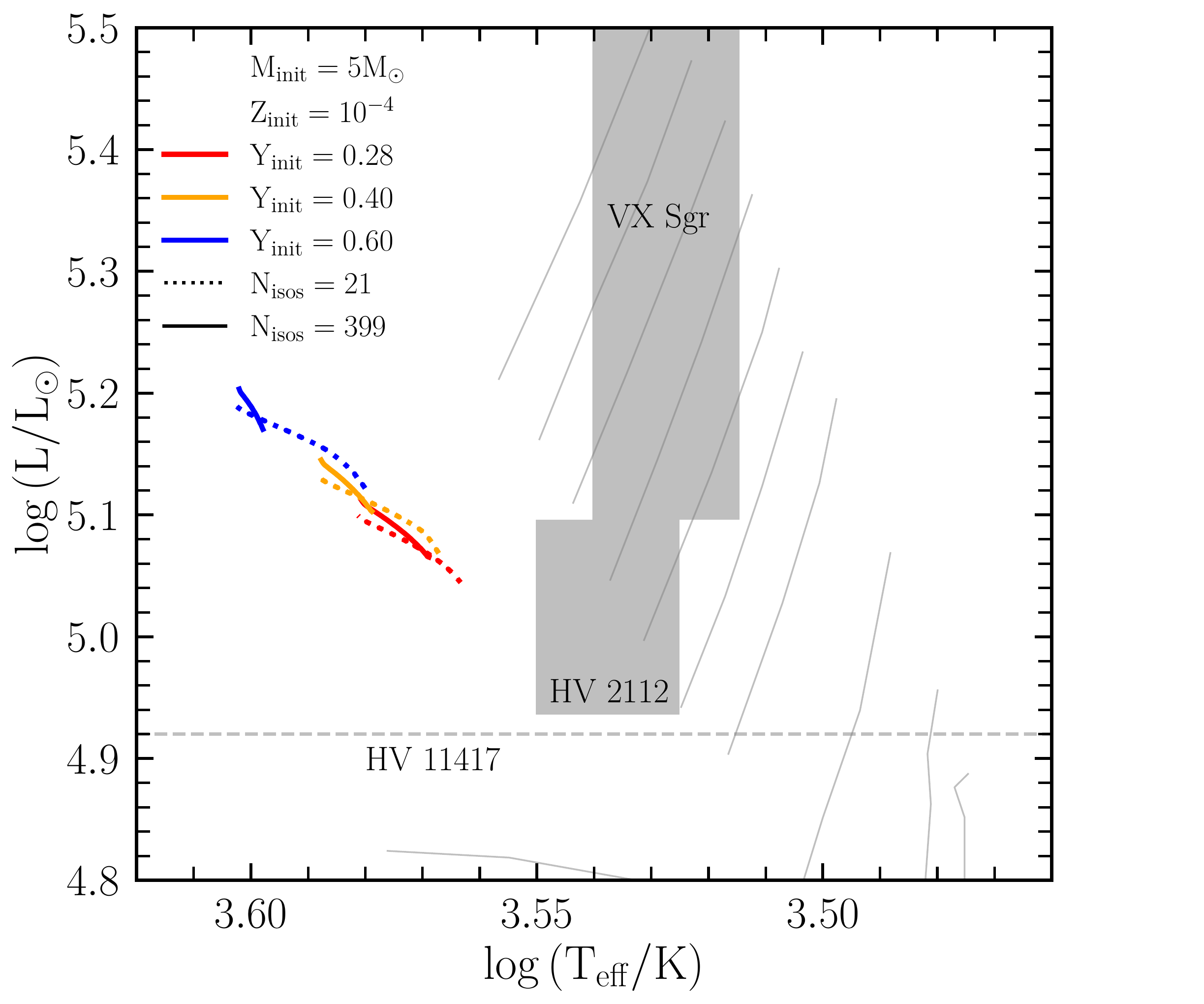}
    \caption{HRDs for variations in the size of the nuclear network. Grey lines and grey boxes have the same meaning as in Figure~\ref{fig:hrd_int_mass}. All models are taken as 5\msun{} with a $\mns=1.4\,\msun$, and use our default parameters. The red lines show  \yinit=0.28, the orange line shows \yinit=0.4, and the blue lines show \yinit=0.6 and are evolved with $\zinit=10^{-4}$.
    Solid lines are models that are evolved with the \nisos{} isotope network, while dotted lines show the 21 isotope network, with the same \yinit. All models were stopped after 25,000 years.}
    \label{fig:hr_nets}
\end{figure}

Figure~\ref{fig:hr_params} shows the HRDs for the parameter variations we have tested here, while Figure~\ref{fig:hr_nets} shows the HRD for a hydrogen-rich and helium-rich \tzo{} evolved with our \nisos{} isotope network.

\subsection{Wind mass loss}

Figure~\ref{fig:hr_wind} shows variations in the wind mass loss efficiency \etanl, which scales the wind mass loss rates of \citet{vanloon05}. All models start with a similar evolution, however as the mass loss rate increases \tzos{} become less luminous for a given temperature. As \etanl{} increases models stop their evolution at higher surface temperatures, live for less time overall, and enrich the surface with fewer metals (at a given age). The effect of wind mass loss on a stellar model is more complex than a single scaling factor \citep{renzo17}.

\subsection{Mixing length alpha}

Figure~\ref{fig:hr_mlt} shows variations in the mixing length \mlta. As  \mlta{} increases the tracks are more luminous and hotter, acting like a more massive \tzo{} (See Figure~\ref{fig:hrd_int_mass}). This increase in luminosity is due to the knee temperature increasing with increasing \mlta{}. Models with $\mlta<1.5$ act similar to the high metallicity models in Section~\ref{sec:comp}. That is, they have a high opacity at the knee, which leads to a low \lknee{}, and thus a low knee temperature.
As \mlta{} increases, \tzos{} produce more metals (at a given age). The age of the \tzo{} is approximately independent of the assumed \mlta.

\subsection{Core mass}

Figure~\ref{fig:hr_mns} shows variations in the neutron star mass $\mns$. As $\mns$ increases stars are more luminous, though at a similar surface temperature. As the \tzo{} evolves, low mass NSs have both a larger change in luminosity and a drop to a lower final luminosity. This is due to the reduction in the lower \ledd{} needed to support the envelope. This behaviour is also related to the ratio of the NS mass to the total mass of the \tzo{}. Higher initial mass \tzos{} show a similar trend as the NS mass (and thus the ratio) is varied. As the mass of the NS increases, the lifetime of the \tzo{} decreases and the surface metal enrichment decreases.

If HV 11417 was a \tzo{} then it is likely to need a very low mass NS $\mns \lesssim 1.2\,\msun$.
At higher core masses (for constant total mass) the knee temperature is lower. This causes a slower metal enrichment in the models and leads to lower production of very heavy elements. Thus if both HV 2112 and HV 11417 were \tzos, we would expect different nucleosynthetic signals.

\subsection{Core accretion rate}

Figure~\ref{fig:hr_ns_acc} shows variations in the accretion efficiency onto the NS $\emedd$. Between $\emedd=10^{-4}$--$10^{2}$, there is no appreciable difference in evolution. When
$\emedd=1.0$ typical accretion rates are $\approx 10^{-9}$--$10^{-8}\,\msunyr$. The $\minit=5\msun$ model has a lifetime of $\approx50,000$ years thus the NS can only gain $\approx 10^{-4}\,\msun$. Only with accretion rates $\emedd>10^3$ does the evolution change significantly, due to the increase in the NS mass, and follows more closely the evolution seen in \citetalias{cannon92}.
Accretion rates in \citetalias{cannon92} are $\approx 10^{-8}$--$10^{-7}\,\msunyr$. These are 10-100 times larger than our accretion rates but still smaller than what we find is needed to force the \tzo{} to evolve to higher luminosities.

\subsection{Nuclear network composition}

Figure~\ref{fig:hr_nets} shows the HRD for our models with both our 21 and \nisos{} isotope nuclear network, for $\minit=5\msun$. We can see that the evolution is broadly similar, with the large nuclear network models being slightly more luminous at the start, but showing a faster decline in their luminosity with time. The largest difference is in the $\yinit=0.6$ models where the large network model evolves in the HRD much less than the \texttt{approx21.net} for the same amount of evolutionary time.  This shows that using an \texttt{approx21.net} is a reasonable assumption to make when considering the bulk properties of a \tzo, especially when considering the significant computational savings for not needing to evolve with a large nuclear network. But this limits the nucleosynthesis signals that an be inferred from the models, without further post-processing.

\section{Network choices}\label{sec:other_nets}

\begin{figure}
    \centering
    \includegraphics[width=\linewidth]{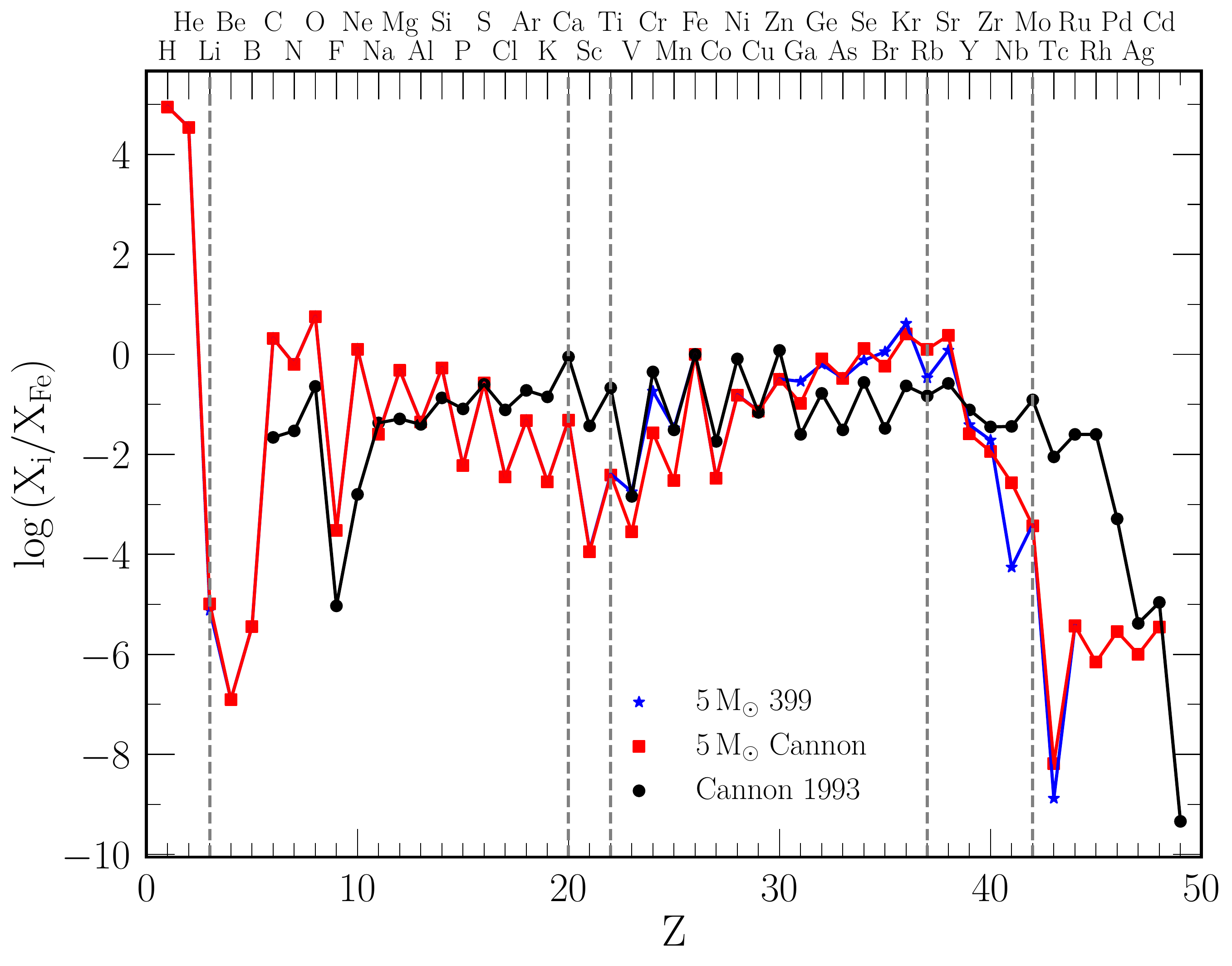}
    \caption{The surface composition relative to solar of a 5\msun{}\tzo{} our default model assumptions (blue star) and a 5\msun{} model with a Cannon-like network (red square).
    This snapshot was taken at $\approx10,000$ years post \tzo{} formation.
    In black dots, the Model A of \citetalias{cannon93}.
    Vertical lines mark elements that may be useful for detecting \tzos.}
    \label{fig:big_net_cannon}
\end{figure}

We have also run an additional network that is the one shown in figure 17 of \citetalias{cannon93}, except we truncate at Cd, leaving 390 isotopes.  Figure~\ref{fig:big_net_cannon} shows the surface mass fractions of our 5\msun{} default model and a 5\msun{} model using the Cannon-like nuclear network. There is very little difference in the overall mass fractions except at V, Cr, Mn, and Nb.

By extending the network to Cd we see a sharp decline in the relative mass fraction of elements heavier than Pd in model A of \citetalias{cannon93}. This is similar behaviour to what is seen in our models, albeit at a higher atomic number. Thus it appears that our models differ mostly by having the element with the highest production factor (Kr) at lower Z, compared to \citetalias{cannon93}. This turn over to lower production factors then also occurs at lower atomic numbers.

\section{MESA physics options}\label{sec:mesa_other}

Radiative opacities are primarily from OPAL \citep{Iglesias1993,
Iglesias1996}, with low-temperature data from \citet{Ferguson2005}
and the high-temperature, Compton-scattering-dominated regime by
\citet{poutanen17}.  Electron conduction opacities are from
\citet{Cassisi2007}.

Nuclear reaction rates are a combination of rates from
NACRE \citep{Angulo1999}, JINA REACLIB \citep{Cyburt2010}, plus
additional tabulated weak reaction rates \citep{Fuller1985, Oda1994,
Langanke2000}. Screening
is included via the prescription of \citet{Chugunov2007}. Thermal
neutrino loss rates are from \citet{Itoh1996}.

\section{Surface mass fractions}

Table \ref{tab:surf_abun} shows the surface chemical mass fractions from Figure~\ref{fig:big_net_y}. These are for a 5\msun{} \tzo{}, with a $1.4\,\msun$ NS, and varying initial helium mass fraction. Values are quoted at 10,000 years after formation. The time evolution of the isotopic data is available in the Zenodo online material.

\begin{table}
\caption{Surface elemental mass fractions for the \nisos{} isotope models with different \yinit{} values. mass fractions measured at 10,000 years post \tzo{} formation. }
\label{tab:surf_abun}
\begin{tabular}{lccc}
\hline
Element & $\yinit=0.28$ & $\yinit=0.40$ & $\yinit=0.60$ \\
\hline
H & 7.194E-01 & 5.993E-01 & 3.991E-01 \\
He & 2.804E-01 & 4.005E-01 & 6.008E-01 \\
Li & 6.087E-11 & 6.105E-11 & 6.127E-11 \\
Be & 1.007E-12 & 1.010E-12 & 1.013E-12 \\
B & 2.886E-11 & 2.894E-11 & 2.903E-11 \\
C & 1.694E-05 & 1.701E-05 & 1.714E-05 \\
N & 5.115E-06 & 5.220E-06 & 5.422E-06 \\
O & 4.605E-05 & 4.617E-05 & 4.629E-05 \\
F & 2.443E-09 & 2.449E-09 & 2.456E-09 \\
Ne & 1.024E-05 & 1.026E-05 & 1.029E-05 \\
Na & 2.045E-07 & 2.050E-07 & 2.056E-07 \\
Mg & 3.883E-06 & 3.893E-06 & 3.904E-06 \\
Al & 3.552E-07 & 3.561E-07 & 3.571E-07 \\
Si & 4.328E-06 & 4.338E-06 & 4.350E-06 \\
P & 4.793E-08 & 4.805E-08 & 4.818E-08 \\
S & 2.160E-06 & 2.165E-06 & 2.171E-06 \\
Cl & 2.869E-08 & 2.875E-08 & 2.888E-08 \\
Ar & 3.850E-07 & 3.859E-07 & 3.871E-07 \\
K & 2.283E-08 & 2.288E-08 & 2.317E-08 \\
Ca & 3.934E-07 & 4.092E-07 & 4.990E-07 \\
Sc & 9.772E-10 & 2.419E-09 & 1.074E-08 \\
Ti & 3.252E-08 & 6.066E-08 & 2.027E-07 \\
V & 1.423E-08 & 3.899E-08 & 1.462E-07 \\
Cr & 1.467E-06 & 1.330E-06 & 7.577E-07 \\
Mn & 2.691E-07 & 4.975E-07 & 1.459E-06 \\
Fe & 8.010E-06 & 8.932E-06 & 1.181E-05 \\
Co & 2.666E-08 & 4.457E-08 & 1.266E-07 \\
Ni & 1.244E-06 & 3.187E-06 & 9.547E-06 \\
Cu & 5.966E-07 & 1.795E-06 & 3.881E-06 \\
Zn & 2.584E-06 & 7.746E-06 & 7.375E-06 \\
Ga & 2.307E-06 & 5.217E-06 & 1.428E-06 \\
Ge & 5.077E-06 & 1.017E-05 & 1.689E-06 \\
As & 2.688E-06 & 4.337E-06 & 3.025E-08 \\
Se & 6.139E-06 & 1.054E-05 & 1.111E-07 \\
Br & 9.040E-06 & 6.373E-06 & 3.187E-09 \\
Kr & 3.322E-05 & 7.479E-06 & 1.132E-09 \\
Rb & 2.707E-06 & 9.555E-09 & 9.490E-11 \\
Sr & 9.650E-06 & 2.832E-08 & 3.171E-10 \\
Y & 3.129E-07 & 7.296E-11 & 6.551E-11 \\
Zr & 1.542E-07 & 1.595E-10 & 1.589E-10 \\
Nb & 4.436E-10 & 1.014E-11 & 1.012E-11 \\
Mo & 3.195E-09 & 3.964E-11 & 3.963E-11 \\
Tc & 1.056E-14 & 1.274E-15 & 6.608E-17 \\
Ru & 2.988E-11 & 2.988E-11 & 2.988E-11 \\
\hline
\end{tabular}
\end{table}


\bsp	
\label{lastpage}
\end{document}